\documentclass[fleqn,usenatbib]{mnras}

\usepackage{newtxtext,newtxmath}

\usepackage[T1]{fontenc}

\DeclareRobustCommand{\VAN}[3]{#2}
\let\VANthebibliography\thebibliography
\def\thebibliography{\DeclareRobustCommand{\VAN}[3]{##3}\VANthebibliography}

\usepackage{graphicx}	
\usepackage{amsmath}	
\usepackage{cleveref}
\usepackage{algorithm}
\usepackage{algpseudocode}
\usepackage{multirow}
\usepackage{bm}

\newcommand\numberthis{\addtocounter{equation}{1}\tag{\theequation}}
\newcommand\codename{{\scriptsize SHIMMERR}}
\newcommand{\mathmat}[1]{\boldsymbol{\mathsf{#1}}}
\newcommand{\mathvec}[1]{\bm{#1}}

\title[Robust DD-calibration far from the target]{Robust direction-dependent gain-calibration of beam-modelling errors far from the target field}
\author[S.A. Brackenhoff et al.]{
S. A. Brackenhoff,$^{1}$\thanks{E-mail: brackenhoff@astro.rug.nl (SAB)}
A.R. Offringa,$^{2,1}$ 
M. Mevius,$^{2}$
L.V.E. Koopmans,$^{1}$
J.K. Chege,$^{1,2}$
E. Ceccotti,$^{1,3}$
\newauthor
C. H\"{o}fer,$^1$
L. Gao,$^{1,4}$
S. Ghosh,$^1$
F.G. Mertens,$^{5,1}$
and S. Munshi$^1$
\\
$^{1}$Kapteyn Astronomical Institute, University of Groningen, PO Box 800, NL-9700 AV Groningen, The Netherlands\\
$^{2}$Netherlands Institute for Radio Astronomy (ASTRON), PO Box 2, NL-7990 AA Dwingeloo, The Netherlands \\
$^3$INAF – Istituto di Radioastronomia, Via P. Gobetti 101, 40129 Bologna, Italy \\
$^{4}$ Liaoning Key Laboratory of Cosmology and Astrophysics, College of Sciences, Northeastern University, Shenyang 110819, China \\
$^5$ LUX, Observatoire de Paris, PSL Research University, CNRS, Sorbonne Université, F-75014 Paris, France
}

\date{Accepted XXX. Received YYY; in original form ZZZ}

\pubyear{\the\year{}}

\begin{document}
\label{firstpage}
\pagerange{\pageref{firstpage}--\pageref{lastpage}}
\maketitle

\begin{abstract}
Many astronomical questions require deep, wide-field observations at low radio frequencies. Phased arrays like LOFAR and SKA-low are designed for this, but have inherently unstable element gains, leading to time, frequency and direction-dependent gain errors. Precise direction-dependent calibration of observations is therefore key to reaching the highest possible dynamic range. Many tools for direction-dependent calibration utilise sky and beam models to infer gains. However, these calibration tools struggle with precision calibration for relatively bright (e.g. A-team) sources far from the beam centre. Therefore, the point-spread-function of these sources can potentially obscure a faint signal of interest. We show that, and why, the assumption of a smooth gain solution per station fails for realistic radio interferometers, and how this affects gain-calibration results. Subsequently, we introduce an improvement for smooth spectral gain constraints for direction-dependent gain-calibration algorithms, in which the level of regularisation is weighted by the expected station response to the sky model. We test this method using direction-dependent calibration method DDECal and physically-motivated beam modelling errors for LOFAR-HBA stations. The new method outperforms the standard method for various calibration settings near nulls in the beam, and matches the standard inverse-variance-weighted method's performance for the remainder of the data. The proposed method is especially effective for short baselines, both in visibility and image space. Improved direction-dependent gain-calibration is critical for future high-precision SKA-low observations, where higher sensitivity, increased antenna beam complexity, and mutual coupling call for better off-axis source subtraction, which may not be achieved through improved beam models alone.
\end{abstract}




\section{Introduction}\label{sec:intro}
To accurately characterise weak astronomical signals in large, complex data sets, it is imperative to separate these signals from radiation originating from other astronomical or terrestrial sources. Such contaminants can be man-made sources of interference, or other sky sources that emit radiation in similar parts of the electromagnetic spectrum. Advanced signal separation methods are needed, because any leakage or confusion may leave the signal of interest obscured. To filter the interfering signals out of the data, they must be well-characterised.

This makes identifying and characterising known sources in the radio sky an important endeavour in sky model-based radio-interferometric calibration. Many catalogues of radio sources that can be used to build a sky model for the field of interest have been constructed, such as the Cambridge surveys \citep{edge_survey_1959,pilkington_survey_1965,baldwin_6c_1985, hales_final_2007}, the WEsterbork Northern Sky Survey (WENSS, \citealt{rengelink_westerbork_1997}), the NRAO VLA Sky Survey (NVSS, \citealt{condon_nrao_1998}), the VLA Low-frequency Sky Survey (VLSS, \citealt{cohen_vla_2007}), the TIFR GMRT Sky Survey (TGSS, \citealt{intema_gmrt_2017}), the GaLactic and Extragalactic All-sky MWA survey (GLEAM, \citealt{hurley-walker_galactic_2017}), the LOFAR Two-metre Sky Survey (LoTSS, \citealt{shimwell_lofar_2017, shimwell_lofar_2022}), and the LOFAR LBA Sky Survey (LoLSS, \citealt{de_gasperin_lofar_2023}).

However, knowledge of the on-sky sources alone is not sufficient to properly remove them from the data, because propagation and instrumental effects affect their signature in the data. For example, ionospheric effects can cause the sources to scintillate \citep{cronyn_interferometer_1972,koopmans_ionospheric_2010}. Furthermore, the chromatic components of the instrument itself, such as the bandpass filter in LOFAR, may cause imprints on all data \citep{de_gasperin_systematic_2019}. Calibration attempts to correct for these effects. For wide fields of view, the most important are the temporally and spatially varying ionospheric and primary beam effects \citep{lonsdale_configuration_2005,smirnov_revisiting_2011}. The two main approaches to removing these are demixing \citep{van_der_tol_self-calibration_2007} and direction-dependent (DD) calibration. Although demixing (i.e. self-calibration on visibilities that have been phase-rotated to the source that needs to be removed) can be effective for removing very bright off-axis sources, it is not as versatile as DD-gain calibration. Demixing is only able to calibrate for a single direction at a time, and the source must dominate the field around it. This places constraints on the minimum source brightness and the minimum separation from other (bright) sources in the vicinity. DD-gain calibration, on the other hand, can cover a larger range of signal-to-noise ratios and source separations, but is computationally more expensive. In DD-gain calibration, different gains are calculated in the directions of groups of sources, rather than for the full sky. This allows for the necessary flexibility to account for DD errors, but also increases the number of degrees of freedom in the calibration process, which creates a risk of overfitting in the gain solutions \citep{patil_systematic_2016, ewall-wice_impact_2017,mourisardarabadi_quantifying_2019}. Such an overfit can lead to suppression of the signal of interest. Therefore, additional constraints have to be introduced to regularise the calibration solutions. Because both propagation effects and individual receiver gains are physically expected to be smooth over a large range of frequencies, one of the most commonly imposed constraints is spectral smoothness. Spectral smoothness is a permissible assumption for most ionospheric effects, but can falter in areas where the instrument primary beam is highly spectrally irregular \citep{trott_spectral_2016}. Such spectral irregularities occur in phased array stations that consist of several antenna elements that are coherently summed to create a larger station. 

Typically, instruments are designed to have a beam that is spatially smooth near the target direction, but this is not necessarily the case outside the main lobe of the primary beam. Especially for phased arrays, this beam can have a very complex spatial structure, such that even closely spaced parts of the sky experience vastly different levels of beam attenuation. This is especially the case near areas where destructive interference results in sharp rises in attenuation, called beam-nulls. Because of the chromatic nature of the instrument beam, and the rotation of the Earth relative to the sky over time, these attenuations also vary spectrotemporally. As a result, the assumption of smoothness is not well-satisfied in areas where the beam changes rapidly as a function of direction, time and frequency. When strong sidelobes (areas of low beam attenuation outside the main lobe) are present and gain-calibration errors are made due to the incorrectly imposed spectral smoothness constraint, the residuals of ill-separated bright interfering sources can still drown out the signal of interest. This can occur even if these sources are spatially separated from the target direction by an angle several times greater than the width of the main lobe of the primary beam\footnote{This is especially the case for the brightest radio sources sometimes referred to as the A-team, such as Cassiopeia A, Centaurus A, Cygnus A, Fornax A, Hercules A, Hydra A, Pictor A, Taurus A, and Virgo A.} \citep{franzen_154_2016, de_gasperin_systematic_2019,trott_impact_2021,munshi_first_2024}. Low-frequency instruments, which tend to have higher off-axis gains and large fields of view, run an especially high risk of having such sources cross their sidelobes.

One way to restore the adherence to the spectral smoothness assumption is to simply impose a weaker constraint (i.e. to reduce the scales on which the gains are expected to be spectrally smooth. While this can help, it also increases the number of degrees of freedom, leading to an increased risk of overfitting, which the smoothness constraint was introduced to prevent. 

Another possibility is to calibrate with an improved model for the beam, such that the most rapid variations are already built into the beam model rather than absorbed into the gains (e.g. \citealt{line_situ_2018,iheanetu_primary_2019,nunhokee_measuring_2020}). This is challenging for phased arrays, however, as their beams change as a function of pointing direction, such that models with many parameters are needed to describe all possible pointings. Most interferometric beam models used in calibration therefore assume identical elements. The station beam is computed by using a full electromagnetic simulation of an element beam, and imposing the station structure through a sum of progressive phase shifts (the `array factor'), ignoring variations in mutual coupling between antenna elements in the array \citep{wijnholds_using_2019}. This type of model describes the general behaviour of the beam well, but suffers from possible instability. If the station has a regular structure, a beam model created with identical elements has deep nulls and strong sidelobes, that may shift or disappear when the station symmetry is broken. This can occur when there are variations in the element beam patterns (e.g. due to mutual coupling), when there are gain variations between elements, or when elements break. Broken symmetries have a smoothing effect on the beam pattern. When the beam model used in calibration is less smooth than the true beam, calibration can imprint the structure of the model beam onto the data.

Models that do incorporate mutual coupling effects are under construction \citep{ohara_uncovering_2024}, but full numerical electromagnetic simulations are not feasible to the required accuracy, because the in situ behaviour of the individual antenna elements is difficult to predict (e.g. \citealt{wilensky_high-dimensional_2024}). Also, failures or degradations within the station may occur. These are not straightforward to detect post-commissioning \citep{chokshi_necessity_2024}. Another option for creating improved beam models is holography using a bright source. One can introduce an artificial bright source (e.g. drone). This is often impossible, however, both because of the many holographic measurements that would be needed to cover the range of possible pointing directions, and because of the distance to the far field. This far field can start kilometres up in the air due to the station sizes, such that flying a sufficiently bright calibrator source over the instrument is not feasible \citep{jacobs_first_2017}. Using satellite carrier emission or reflections instead is a possibility, but these signals are typically narrow-band and therefore cannot be used to create an accurate model across the full spectral band of the instrument. Finally, bright sky-sources can be used, but only in a limited set of directions \citep{nunhokee_measuring_2020}. To create a model across the full band and for all pointings, most modern radio observatories opt for simplified beam models instead. However, these falter in the presence of real-world effects, such as broken antenna elements, drifting or degraded amplifiers, or electromagnetic coupling within the station. Although these models are useful to first order and sufficient in many cases, they do leave errors which are prohibitive in the most ambitious science cases. 

Errors in the beam model are an issue for all interferometers, but they are most prominent for regular phased arrays, where any gain error can break the symmetry within a station and therefore change the shape of the primary beam not just quantitatively, but also qualitatively. This is the case for the stations of instruments such as the High-Band Antennas of the Low Frequency Array (LOFAR-HBA, \citealt{van_haarlem_lofar_2013}), the New Extension in Nan\c{c}ay Upgrading LOFAR (NenuFAR, \citealt{zarka_lssnenufar_2012}), and the Murchison Widefield Array (MWA, \citealt{bowman_science_2013}). Furthermore, it is also a prominent issue for interferometers with complex wide-field antenna beams, such as the upcoming low band part of the Square Kilometre Array (SKA-Low, \citealt{dewdney_square_2009}). Beam modelling errors affect a wide range of high dynamic range science cases. For example, transient surveys \citep{murphy_vast_2013}, 
the observation of the faint large-scale structure traced out by neutral hydrogen \citep{hale_lofar_2019}, or the creation of deep extragalactic surveys \citep{best_lofar_2023}. There is also the continuing effort to detect the redshifted 21-cm signal from the Epoch of Reionization (EoR) and Cosmic Dawn with the aforementioned instruments (e.g. \citealt{koopmans_cosmic_2015,trott_deep_2020, munshi_first_2024,mertens_deeper_2025}).

We illustrate the effect of DD beam-modelling errors on the gains recovered during DD gain-calibration using the EoR science case and the LOFAR-HBA instrument, because this example is especially affected by DD errors \citep{patil_systematic_2016,ewall-wice_impact_2017,sardarabadi_identifiability_2019}. These errors are likely one of the dominant causes of unwanted residual emission ("excess variance") seen in the data after sky model subtraction with the DD-gain solutions applied \citep{gan_statistical_2022,mertens_deeper_2025}. Beam errors have also been shown to be limiting this science case on similar instruments such as the MWA \citep{chokshi_necessity_2024} and NenuFAR \citep{munshi_first_2024}. Because the short baselines that are the main focus of interferometric EoR science are not strongly affected by the ionosphere \citep{brackenhoff_ionospheric_2024}, we can focus mainly on the beam itself. 

We present the impact of realistic beam errors on LOFAR-HBA with a Python code named \codename\ (Station Heterogeneity Impact on Multi-dimensional beam-Modelling Errors simulatoR and calibratoR)\footnote{\href{https://github.com/Stefanie-B/shimmerr}{https://github.com/Stefanie-B/shimmerr}}. We create beam errors by simulating that a number of individual antenna elements within the station are switched off, as this is expected to be a dominant source of beam modelling errors and differences between individual LOFAR-HBA stations. We calculate the resulting beam changes, the effects they have on the visibility data, and the impact on the spectrally regularised calibration of sources far from the target direction. This is the first simulation that shows the full end-to-end effect of beam errors on LOFAR-HBA DD-calibration on the scale of a full night of observations. Furthermore, we present a first potential heuristic solution to the broken spectral smoothness assumption in DD-gain calibration and show that this solution improves upon the standard method both in visibility and image space. In this more robust regularisation method, the regulariser is weighted by the forward prediction of the visibility contribution of each direction, such that time- and frequency intervals with strong beam attenuation do not negatively affect the gain solutions at other times and frequencies. We use the observation and calibration settings from LOFAR EoR to illustrate the impact of beam errors, but emphasize that these results apply to many science cases and instruments, especially those needing the shorter baselines.

This paper is structured as follows: the details of DD-gain calibration and the errors introduced by incorrectly assuming spectral smoothness are explained in \Cref{sec:problem}. We also elaborate on the heuristic to mitigate this issue here. In \Cref{sec:sims}, the simulation and calibration setups are explained. The results of the calibration are discussed in \Cref{sec:results}, where we first discuss how the standard and proposed method impact the obtained gains. We then examine how errors in these gains manifest in visibility space, and finally, we assess their impact on images. We focus on the shortest baselines in particular, as this type of DD-calibration is uniquely equipped to calibrate these. We conclude the paper in \Cref{sec:discussion}.

\textit{Notation:} Throughout this work, we indicate vectors with lowercase bold italic typesetting ($\mathvec{a}$), unit-length vectors with a hat ($\mathvec{\hat{a}}$) and matrices with uppercase bold sans-serif typesetting ($\mathmat{A}$). The complex conjugate, complex conjugate transpose and matrix inverse are indicated by $(\cdot)^*$, $(\cdot)^H$, and $(\cdot)^{-1}$, respectively. Averages are denoted by $\left<\cdot\right>$, the Frobenius norm (the square root of the sum of squares of the elements of a matrix) is denoted by $\|\cdot\|_F$. The inner product is denoted with $\cdot$, the Hadamard product by $\odot$ and the imaginary unit by $i$.
\section{Direction-dependent gain-calibration outside the main beam lobe} \label{sec:problem}
In radio astronomy, the sky is described by its brightness distribution, denoted by $I(l,m)$, where $l$ and $m$ are direction cosines. $I(l,m)$ represents the specific intensity as a function of sky direction. Under the assumptions of spatial incoherence of $I(l,m)$ and quasi-monochromatic radiation, the van Cittert-Zernike theorem provides the theoretical basis for radio interferometry. The theorem states that the spatial coherence of the electric field (i.e. a visibility) is given by the Fourier transform of the sky brightness distribution $I$ \citep{thompson_interferometry_2001}. A baseline (i.e. the vector connecting two receiver elements) with components $(u,v,w)$ in units of wavelength measures a visibility $V(u,v,w)$,
\begin{align}
     V(u,&\varv,\varw)=\iint A(l,m)I(l,m) 
     \label{eq:VCZ}\\
     &\exp\left\{-2\pi i\left[ul+\varv m+\varw\left(\sqrt{1-l^2-m^2}-1\right)\right]\right\}\frac{dldm}{\sqrt{1-l^2-m^2}}\nonumber.
\end{align}
Here, $l,m$ are direction cosines on the sky, and $A(l,m)$ is the beam response of the receiver. 

Whilst the van Cittert-Zernike theorem is an excellent tool for understanding the physical meaning of visibilities, it has its limitations, because it can only describe scalar fields (i.e. no polarisation effects), and does not include instrumental or propagation effects. The Van-Cittert Zernike theorem can be generalised to the Radio Interferometric Measurement Equation (RIME, \citealt{hamaker_understanding_1996,smirnov_revisiting_2011-1}). For a single point-source, the RIME is given by
\begin{equation}
    \mathmat{V}_{p,q}(\nu,t)=\mathmat{P}_p(\nu,t) \mathmat{C}_{p,q}(\nu,t) \mathmat{P}_q^{H}(\nu,t) + \mathmat{N}_{p,q}(\nu,t) .
\end{equation}
\noindent In this equation, $ \mathmat{V}_{p,q}$ is a $2\times 2$ matrix that describes the visibility on the baseline between receiver elements $p$ and $q$ at time $t$ and frequency $\nu$ with the Jones formalism. $\mathmat{P}_p$ is a $2\times2$ propagation matrix describing the signal path of incident radiation on receiver element $p$, as well as instrumental effects. Finally, $\mathmat{N}_{p,q}$ describes white Gaussian noise\footnote{Note that the elements of $\mathmat{N}$ describe different polarisations and not the statistical behaviour of the noise}. $\mathmat{C}_{p,q}$ is the $2\times2$ source coherency matrix, which describes the correlations between orthogonal components of the electric field originating from the source. Let $\mathvec{e}=\left[e_x, e_y\right]^T$ be the complex Jones vector that represent the complex electric field of the source. The source coherency matrix is then defined as
\begin{equation}
    \mathmat{C}_{p,q}(\nu,t)=\left<\mathvec{e}\mathvec{e}^H\right>,
\end{equation}
where the averages $\left<\cdot\right>$ are over time. This matrix describes both the total intensity and the polarisation of incoming radiation.

For uncorrelated, discrete sources, the brightness of the full sky is additive, because the electric fields from incoherent sources sum in power rather than amplitude. If we assume all components are sufficiently compact that the Jones matrix can be assumed constant over the extent of the source, the full-sky RIME can be broken up into a set of components denoted by $d$, i.e. 
\begin{equation}
    \mathmat{V}_{p,q}(\nu,t)= \sum_d \mathmat{P}_{p,d}(\nu,t)\mathmat{C}_{p,q,d}(\nu,t) \mathmat{P}_{q,d}(\nu,t)^{H} + \mathmat{N}_{p,q}(\nu,t).
    \label{eq:fullsky_RIME}
\end{equation}

The RIME is a very powerful tool for describing the process of calibration. In sky model-based calibration, a sky model is created to formulate $\mathmat{C}_{p,q,d}$. This model typically consists of a set of discrete sources that are forward modelled to describe their coherency as a function of baseline, space and time to obtain the set of brightness coherency matrices. If a model of the primary beam $\mathmat{A}_{p,d}$ is known, it is applied to the sky-model during forward modelling.

In standard radio interferometry, individual antennas or dishes are used as receiver elements (i.e. directly correlated to form visibilities). However, the receiver elements in many modern interferometers, such as LOFAR, SKA-Low, NenuFAR, and MWA, are phased array stations, rather than single antennas. Each station consists of a set of $N_m$ antennas that are coherently summed to create electronically steerable beams. By introducing phase shifts based on the antenna element positions $\mathvec{x}_{p,m}$ in metres, the station can be steered in direction $\mathvec{\hat{s}_0}$,
\begin{equation}
    \mathvec{\varv}_p(\nu,t)=\sum_m \exp\left[-\frac{2\pi i\nu}{c}\mathvec{\hat{s}_0}\cdot\mathvec{x}_{p,m}\right]\mathvec{\varv}_{p,m}(\nu,t).
\end{equation}
Here, $c$ is the speed of light in metres per second. $\mathvec{\varv}_p(\nu,t)$ is the $2\times 1$ beamformed voltage vector of station $p$, whereas $\mathvec{\varv}_{p,m}(\nu,t)$ is the voltage from the $m$-th element in the station. The elements of the voltage vectors describe the orthogonal polarisation feeds. The visibility in \Cref{eq:fullsky_RIME} is given by the correlation between two stations,
\begin{equation}
    \mathmat{V}_{p,q}(\nu,t)=\mathvec{\varv}_p(\nu,t)\mathvec{\varv}^H_q(\nu,t).
\end{equation}

A phased array can determine the angle of arrival from the delay by which a wavefront is received by its elements. A wavefront originating from a source in direction $\mathvec{\hat{s}}$ has a wave vector of $\mathvec{k}(\mathvec{\hat{s}}, \nu)=-\frac{2\pi\nu}{c}\mathvec{\hat{s}}$~[m$^{-1}$]. These delays can be written as $\exp\left(-i\mathvec{k}\cdot\mathvec{x}_{p,m}\right)$. The station response to a source can be modelled as
\begin{equation}
\mathmat{A}_p( \mathvec{\hat{s}},\nu)=\mathmat{U}_p(\mathvec{\hat{s}},\nu)\sum_m\mathmat{G}_{p,m}\exp\left[-i\mathvec{k}(\mathvec{\hat{s}}-\mathvec{\hat{s}_0},\nu)\cdot\mathvec{x}_{p,m}\right],
\label{eq:DDECal_A}
\end{equation}
\noindent where $\mathmat{G}_{p,m}$ are per-element complex gains, and $\mathmat{U}_p(\mathvec{\hat{s}},\nu)$ is the antenna element beam (e.g. a simple dipole), which we assume to be identical between antennas here. The element-beam independent part of this equation,
\begin{equation}
\mathrm{AF}(\mathvec{\hat{s}},\nu)=\sum_m\mathmat{G}_{p,m}\exp\left[-i\mathvec{k}(\mathvec{\hat{s}}-\mathvec{\hat{s}_0},\nu)\cdot\mathvec{x}_{p,m}\right],
\label{eq:array_factor}
\end{equation} 
is called the `array factor'. It represents the far-field beam of the station based solely on the element positions and their excitations, assuming identical, isotropic elements. 

The propagation matrix is then split into the beam model, and a model for all other propagation effects (both astronomical and instrumental), i.e. $\mathmat{P}_{p,d}=\mathmat{J}_{p,d}\mathmat{A}_{p,d}$. Calibration then attempts to find the matrices $\mathmat{J}_{p,d}(\nu,t)$ that make this sky model best match the data. Because the sky usually remains coherent over short time and frequency intervals, a matrix $\mathmat{J}_p$ is found over such an interval  instead of per data point, both raising the signal-to-noise ratio and reducing the computational cost. We call these intervals `solution intervals' and define a parameter $l$ that iterates over all combinations of $\nu$ and $t$ within a solution interval to simplify notation. For similar reasons, groups of spatially close sources (calibration clusters, denoted by $k$) are assigned the same Jones matrix $\mathmat{J}_{p,k}$, rather than an individual matrix per source $\mathmat{J}_{p,d}$. This is an approximation, since the clustering of sources is a heuristic procedure based on the required minimum flux level of a calibration cluster, or the total permissible number of calibration clusters. This means that one pair of closely spaced sources can be assigned the same calibration cluster, whereas another similarly spaced source could be assigned a different matrix, even though its propagation path can be very similar to that of the original pair. If the clusters are not well-chosen, the gain can abruptly change on the boundaries between calibration clusters, effectively imposing a two-dimensional step-function on the spatial variations in propagation. The matrix $\mathmat{J}_{p,k}$ models the typical propagation effects for sources in a calibration cluster, but is an approximation for each source individually. The simplest form of the calibration problem, for each solution interval separately, then becomes
\begin{align}
    \underset{\mathmat{J}_{p,k}}{\text{Minimise    }}&\sum_{p,q}\ \sum_{l}\\&\nonumber
    \left\|\mathmat{V}_{p,q,l} - \sum_k\mathmat{J}_{p ,k}\left(\sum_{d\in k}\mathmat{A}_{p,d,l}\mathmat{C}_{p,q,d,l} \mathmat{A}_{q,d,l}^{H}\right)\mathmat{J}_{q,k}^{H}\right\|^2_F.
\end{align}
\noindent The aim is then to absorb the largest imperfections in the (lack of) beam model into the calculated solutions $\mathmat{J}_{p,k}$.

Typically, there are several spectral solution intervals stacked within the bandwidth of the instrument. A separate gain ($\mathmat{J}_{p,k}$) exists for each frequency solution interval $\Delta\nu_{\mathrm{sol}}$. Applying a spectral smoothness constraint means assuming that each of these $\mathmat{J}_{p,k}$ lie on a smooth function of $\nu$. The calibration problem now becomes
\begin{align*}
    \underset{\mathmat{J}_{p,k}}{\text{Minimise    }}&\sum_{\nu_{\mathrm{sol}}}\sum_{p,q}\ \sum_{l}\\
    \Bigg\|\mathmat{V}_{p,q,l} - &\sum_k\mathmat{J}_{p ,k}(\nu_{\mathrm{sol}})\left(\sum_{d\in k}\mathmat{A}_{p,d,l}\mathmat{C}_{p,q,d,l} \mathmat{A}_{q,d,l}^{H}\right)\mathmat{J}_{q,k}^{H}(\nu_{\mathrm{sol}})\Bigg\|^2_F\\ \\
    \text{subject to   }&\hspace{0.1cm}\text{spectral smoothness in }\mathmat{J}_{p,k}(\nu_{\mathrm{sol}}) .\numberthis
    \label{eq:DDcal_optimizationproblem}
\end{align*}
\noindent This is a non-convex fourth-order complex optimization problem. There is no analytical solution, which means that careful heuristic choices need to be made to obtain an numerical approximate solution. 

There have been many efforts to create DD-gain calibration methods. Examples are joint calibration and imaging algorithms \citep{repetti_non-convex_2017,arras_unified_2019,birdi_polca_2020,roth_bayesian_2023}, or demixing, in which each cluster of sky-model components is subtracted separately through direction-independent self-calibration on phase-rotated visibilities \citep{van_der_tol_self-calibration_2007}. This latter approach suffers from a high number of degrees of freedom, such that it can only be applied for the brightest sources and at large separations. It is not able to remove several bright sources at a relatively large separation in the main beam, or faint sources in the sidelobes, for example. A modified approach, in which all other sources are subtracted with the (imperfect) calibration-gains during each direction-independent calibration step, called peeling, has also been widely used \citep{van_weeren_lofar_2016,albert_probabilistic_2020}. In this work, however, we consider the families of calibration algorithms that simultaneously calibrate all directions during each iteration, and implement the smoothness constraint to limit the number of degrees of freedom\footnote{Algorithms such as {\scriptsize FACETCAL} \citep{van_weeren_lofar_2016} also depend on a smooth beam approximation and therefore may suffer from similar effects, however, our solution method is not straightforward to implement in {\scriptsize FACETCAL}.}. 

The two most important algorithms in this context are {\scriptsize SAGECAL-CO} \citep{yatawatta_distributed_2015} and {\scriptsize DDECAL} \citep{gan_assessing_2023}. {\scriptsize SAGECAL-CO} uses consensus optimisation to penalise gains for deviating from functions in a low-order Bernstein polynomial basis. These Bernstein polynomials are optimised during calibration as well. {\scriptsize SAGECAL-CO} therefore does not directly enforce smoothness during each iteration, but does provide a `global collector', which contains a smooth approximation of the obtained gains in this Bernstein basis. This consensus-based approach allows the solutions to be split over frequency channels, and therefore, be solved in a highly parallel manner. {\scriptsize DDECAL} enforces spectral smoothness at each iteration through low-pass filtering the gains with a truncated Gaussian kernel in between iterations. We apply {\scriptsize DDECAL} in this work, because it makes understanding the effects of spectral regularisation with erroneous modelling assumptions more intuitive, and because it is part of the widely used {\scriptsize DP3} processing software\footnote{\href{https://dp3.readthedocs.io/en/stable/}{dp3.readthedocs.io}} \citep{van_diepen_dppp_2018}. However, we emphasize that although {\scriptsize SAGECAL-CO} and {\scriptsize DDECAL} apply the spectral smoothness constraint differently, they both suffer from the same issue; the fact that a spectrally smooth gain function $\mathmat{J}_{p,k}(\nu_{\mathrm{sol}})$ does not accurately describe the spectral variation of beam-attenuation modelling errors, especially outside the main lobe of the primary beam. Our proposed method can be implemented straightforwardly in {\scriptsize SAGECAL-CO} as well as in {\scriptsize DDECAL}. We expect our conclusions to generalise to {\scriptsize SAGECAL-CO}, and other similar DD-gain calibration algorithms that impose spectral smoothness.

\subsection{DDECal}
{\scriptsize DDECAL} contains several algorithms\footnote{These are {\tt directionsolve}, {\tt directioniterative}, {\tt hybrid} and {\tt LBFGS}. The {\tt LBFGS} routine was implemented in {\scriptsize DDECAL} specifically to compare results to those obtained with {\scriptsize SAGECAL-CO} and not widely used in isolation. {\tt directioniterative} and {\tt hybrid} are mainly used to reduce the required computational time, which is why we focus on the more precise {\tt directionsolve} in this work.} that numerically approximate the solution of \Cref{eq:DDcal_optimizationproblem}. Below, we describe its {\tt directionsolve} algorithm, which is similar to the algorithm described by \citet{smirnov_radio_2015}, but with the addition of a constraints framework. The solver finds the gains by solving for one station in all directions simultaneously. For all other stations, the gains of the previous iteration are assumed. In \Cref{eq:DDcal_optimizationproblem}, this means that $\mathmat{J}_{q,k}$ are assumed to be known, while for each $p$, $\mathmat{J}_{p,k}$ are updated for all directions $k$ simultaneously. For each solution interval, the data and gains are stacked column-wise, as\footnote{The notation used here is a corrected version of that presented by \citet{gan_assessing_2023}, but is consistent with a new version of their notation available at \href{https://arxiv.org/abs/2209.07854v2}{https://arxiv.org/abs/2209.07854v2}.}, 
\begin{multline}
    \mathmat{\widetilde{V}}_p = 
    \left[
        \mathmat{V}_{p,q=0,l=0}\hspace{3mm}
        \mathmat{V}_{p,q=0,l=1}\hspace{3mm}
        \dots\hspace{3mm}
        \mathmat{V}_{p,q=0,l=N_l}\hspace{3mm}
        \right. \\ \left.
        \mathmat{V}_{p,q=1,l=0}\hspace{3mm}
        \dots\hspace{3mm}
        \mathmat{V}_{p,q=N_s,l=N_l}
    \right]
    \label{eq:DDECal_V},
\end{multline}
and 
\begin{equation}
    \mathmat{\widetilde{J}}_p = 
    \begin{bmatrix}
        \mathmat{J}_{p,k=0}&\mathmat{J}_{p,k=1}&\dots&\mathmat{J}_{p,k=N_k}
    \end{bmatrix}.
    \label{eq:DDECal_J}
\end{equation}
\noindent Here, $\mathmat{\widetilde{J}}_p$ is the $2\times2N_k$ matrix of gains, where $N_k$ is the number of calibration directions. $\mathmat{\widetilde{V}}_p$ is the $2\times2N_sN_l$ matrix of visibilities associated with station $p$. $N_s$ is the total number of stations that station $p$ shares a baseline with, and $N_l$ are the number of time and frequency steps within a solution interval. 

Using the sky and beam models, and the DD-gains from the previous iteration, a model matrix $\mathmat{\widetilde{M}}_p$ is also constructed. This is a $2N_k\times2N_sN_l$ matrix with its rows stacked in the same order as the columns in \Cref{eq:DDECal_V}, and elements given by
\begin{align}
    \left(\mathmat{\widetilde{M}}_p\right)_{k,(ql)}=\sum_{d\in k}        \mathmat{A}_{p,l,d}\mathmat{C}_{p,q,l,d} \mathmat{A}_{q,l,d}^{H}\mathmat{J}_{q,k}^{H}.
    \label{eq:DDECal_M}
\end{align}
\noindent Using these equations, the DD RIME reduces to 
\begin{equation}
    \mathmat{\widetilde{V}}_p = \mathmat{\widetilde{J}}_p\mathmat{\widetilde{M}}_p.
    \label{eq:DDECal_LS_problem}
\end{equation}
\noindent This is a least-squares fit to the RIME for each station (at each time and frequency solution interval) separately.

Because this is a non-linear least-squares problem (given that the matrices $\mathmat{\widetilde{M}}_p$ depend on $\mathmat{\widetilde{J}}_p$ and will therefore change after $\mathmat{\widetilde{J}}_p$ has been updated), a Gauss-Newton step-size is used for convergence. Denoting the iteration number by $[n]$, a full iteration of {\scriptsize DDECAL} follows the procedure described below.

\begin{enumerate}
    \item \textbf{Prediction:} Compute matrices $\mathmat{\widetilde{M}}_p[n]$ using \Cref{eq:DDECal_M}.
    \item \textbf{Least-squares solution:} We denote the solution to \Cref{eq:DDcal_optimizationproblem} given the known matrices $\mathmat{\widetilde{V}}_p$ and $\mathmat{\widetilde{M}}_p[n]$ as $\mathmat{\widetilde{L}}_p[n+1]$, to differentiate between this intermediate computation step and the new gain $\mathmat{\widetilde{J}}_p[n+1]$ that is obtained after regularisation. This least-squares update is computed for each spectral solution interval $\Delta\nu_{\mathrm{sol}}$.
    \item \textbf{Update the intermediate gains:} To avoid oscillation or divergence through overshooting the global minimum, a Gauss-Newton descent with an update speed of $\alpha=0.2$ is used. A noisy version of the new gain is found using  $\mathmat{\widetilde{U}}_p[n+1]=(1-\alpha)\mathmat{\widetilde{J}}_p[n]+\alpha\mathmat{\widetilde{L}}_p[n+1]$.
    \item \textbf{Regularisation:} Finally, the new gains are obtained by regularising the intermediate gain solutions. Although the {\scriptsize DDECAL} framework also contains other constraints, we focus on spectral regularisation, in which the least-squares fits are low-pass filtered through convolution with a truncated Gaussian kernel. The standard deviation $\sigma$ of this kernel is set by the user to a given spectral smoothness scale, and the kernel is truncated at three times its standard deviation. The wider this filter is, the smoother the obtained gains will be. Each component of the Jones matrix $\mathmat{\widetilde{J}}_p[n+1](\nu_{\mathrm{sol}})$ is separately regularised. For each component, a smooth update is given by
    \begin{equation}
    \widetilde{J}_{p,k}(\nu)=\frac{\sum_{\nu'\in [\nu-3\sigma, \nu+3\sigma]}{K(\nu,\nu')W_{p,k}(\nu')U_{p,k}(\nu')}}{\sum_{\nu'\in [\nu-3\sigma, \nu+3\sigma]}{K(\nu,\nu')W_{p,k}(\nu')}},
    \label{eq:DDECal_smoothing}
    \end{equation}
    where we have dropped the subscript `sol' for legibility. The Gaussian kernel is given by $K(\nu,\nu')=\exp[-\frac{1}{2}(\nu-\nu')^2/\sigma^2]$, and $W_{p,k}(\nu)$ is a weight. Such weights are useful when some solution intervals contain fewer data points than others, due to interference excision, for example\footnote{In this work, no interference is simulated and no such excision is performed, so these weights equal one.}.
\end{enumerate}

\subsection{Direction-dependent gain-calibration with non-smooth beams}
When the beam is rapidly varying as a function of frequency, however, the assumptions underlying \Cref{eq:DDcal_optimizationproblem} break down. To explain this, we define the `true beam', i.e. the station's effective beam with in situ effects (in the case of our simulations, these consist of switched-off antenna elements), and the `model beam' or `calibration beam', which is the expected beam model $\mathmat{A}_{p,l,d}$ (without missing elements). The most obvious issue is that there may be modelling errors between the true beam and the beam model on scales smaller than a solution interval in frequency and time. In this case, the single $2\times2$ Jones matrix $\mathmat{J}_{p,k}$ computed per solution interval and direction cannot describe the small-scale beam modelling errors accurately. However, deviations from the beam model within the station typically lead to the primary beam being smoother than the ideal model beam. This is because beam shape variations, and amplitude and phase deviations between elements in a phased station break the station symmetry. Therefore, they tend to blur sharp features in the beam, but also tend to lead to a less sharp fall-off of the amplitude of the far sidelobes, as we illustrate in \Cref{subsec:sims_brokenelements,subsec:results_gains}. As long as a sufficiently small solution interval is chosen, the errors introduced by these intra-interval modelling errors are expected to be small, because the beam modelling errors do not lead to more rapidly varying errors than are present in the model beam.

\begin{figure}
    \centering
    \includegraphics[width=\linewidth]{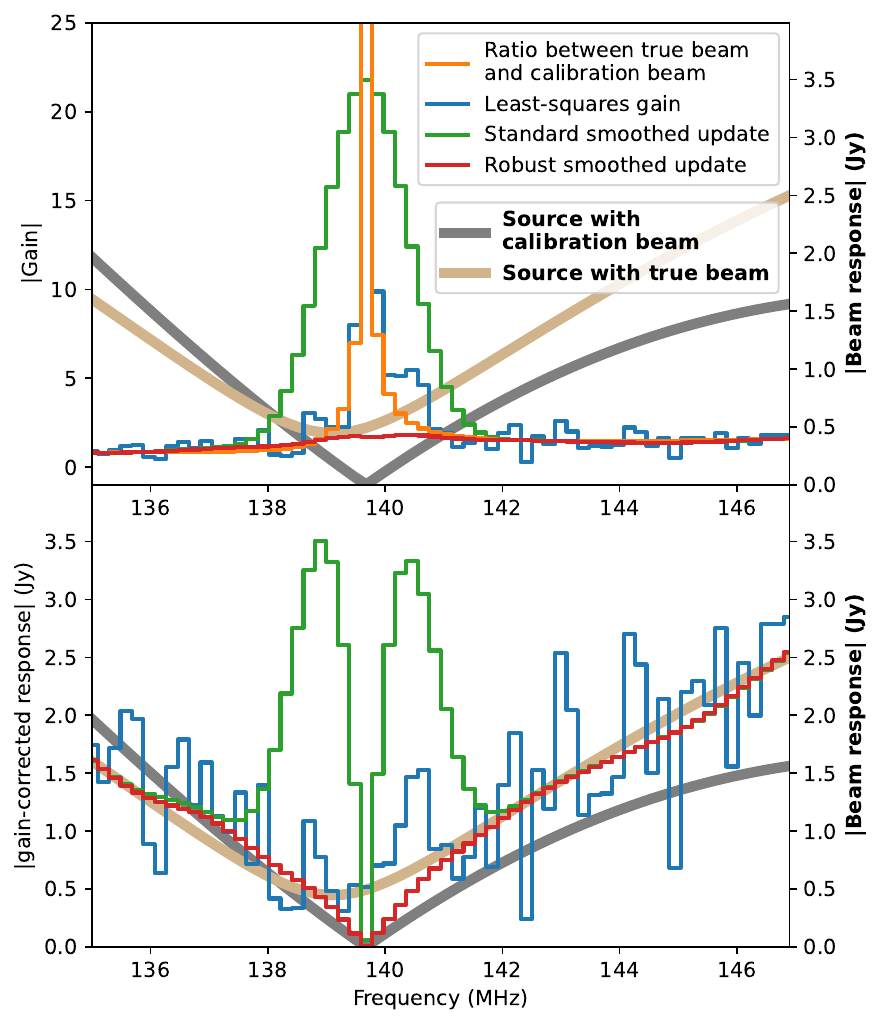}
    \caption{Illustration of the effect of an erroneous beam model on calibration. In both panels, the bold axis shows the prediction of the response of a station to a single point-source, both with the beam model used in calibration (grey) and the actual perturbed beam (tan). Top: The regular weight axis shows the gain solutions for the same station. Here, blue is the unregularised solution $ {J}_{p,k}(\nu_{\mathrm{sol}})$, green is the same solution after standard smoothing has been applied and red shows the same with the proposed robust regularisation. Orange illustrates the ideal gain that would perfectly match the calibration and true beam. Bottom: the regular weight axis shows the product of the source in the calibration beam and the colored lines from the top panel. For perfect calibration, this aligns with the tan line.}
    \label{fig:beam_cartoon}
\end{figure}

The second, bigger problem, and focus of this work, occurs when the beam varies strongly within a spectral smoothness interval. We illustrate this effect for an extreme case with a single polarisation in \Cref{fig:beam_cartoon}. In both panels, the right-hand axis (corresponding to the thick lines) shows how the apparent flux of a single point-source that is far from the target direction varies as a function of frequency. In this example, we use the position of Cas A from the point-source model discussed in \Cref{subsec:sky_model}, but use a flat-spectrum source with an intrinsic flux of 2.1~kJy. Cas A has a separation from the target field of $\sim30^\circ$. The primary beam width of a LOFAR-HBA station is $\sim4^\circ$ in the simulated frequency range, placing the source well into the sidelobes of the station beam. The thick tan-coloured line corresponds to how the source is `seen' through the true primary beam of a single station (this is the beam in which elements have been switched off). We denote this as ${ {A}'}_{\!\!p,l,\mathrm{s}} {I}_\mathrm{s}$, where ${ {A}'}_{\!\!p,l,\mathrm{s}}$ is the true primary beam in the direction of Cas A for a single polarisation, and $ {I}_\mathrm{s}$ is the intrinsic flux of the source. The thick grey line shows the same with an ideal model primary beam (with all elements functioning), which we denote as $ {A}_{p,l,\mathrm{s}} {I}_\mathrm{s}$ (without a prime). The latter has an area of very high attenuation (a `null'), that is spectrally shifted, wider, and less deep in the true beam ${ {A}'}_{\!\!p,l,\mathrm{s}}$. We simulate with a monochromatic antenna beam, such that all spectral effects here stem from beamforming, i.e. \Cref{eq:array_factor}.

The left-hand side axes (in regular weight), show mock calibration results after a single iteration. In the top panel, we illustrate gains $ {J}_{p,\mathrm{s}}$\footnote{Because the calibration direction contains only a point source, we use the same subscript for $k$ and $d$.} that would result from calibration with different types of regularisation. In the bottom panel, these gains are applied to the source's model apparent flux with the calibration beam, i.e. $ {J}_{p,\mathrm{s}} {A}_{p,l,\mathrm{s}} {I}_\mathrm{s}$. Applying the gains for both stations in a baseline creates model visibilities that are corrupted in the same way as the data. These corrupted model visibilities are subtracted from the data to remove the source. Therefore, the calibrated model flux $ {J}_{p,\mathrm{s}} {A}_{p,l,\mathrm{s}} {I}_\mathrm{s}$ shown in regular weight in the bottom panel should ideally coincide with the true response ${ {A}'}_{\!\!p,l,\mathrm{s}} {I}_\mathrm{s}$ shown by the thick tan-coloured line. With our assumptions, the ideal gain (that would calibrate this direction perfectly) is equal to the ratio between the true beam and the calibration beam ${ {A}'}_{\!\!p,l,\mathrm{s}}/{ {A}'}_{\!\!p,l,\mathrm{s}}$. This gain is shown as the orange line in the top panel. Because the true beam ${ {A}'}_{\!\!p,l,\mathrm{s}}$ is unknown, calibration must estimate this ratio from the noisy observations.

In a noisy observation, the first-iteration least-squares solution per spectral solution interval $ {\widetilde{L}}_p$ looks like the blue line in the top panel. When no regularisation is applied, this will be the gain after the first iteration $ {J}_{p,\mathrm{s}}$. There is a very large spike in the gain at the null in calibration beam. Furthermore, due to the many degrees of freedom, the observational noise is still clearly visible in the least-squares solution, as relatively large spectral channel-to-channel variations throughout the band in the blue line in the top panel. This is normally not a desirable solution, because having too many degrees of freedom leads to overfitting and an undesired interplay between the gains computed for different calibration directions. Additionally, applying noisy solutions to a bright sky model leads to increased errors in the residual data when these sources are subtracted. For unregularised calibration, the value of $ {J}_{p,\mathrm{s}} {A}_{p,l,\mathrm{s}} {I}_\mathrm{s}$ (blue line in the bottom panel) therefore does not model the actual source response ${ {A}'}_{\!\!p,l,\mathrm{s}} {I}_\mathrm{s}$  (thick tan-coloured line) well.

When the low-pass filter (i.e. the truncated Gaussian kernel) is applied to constrain the solution, we obtain the green lines. Although the noise on the gain has been reduced, the channels adjacent to the null have a higher gain when regularisation is applied. This does not model the true source well near the null, as is evident from comparing the green and thick tan-coloured lines in the bottom panel of \Cref{fig:beam_cartoon}. In this way, erroneous gains obtained in the null can spread to channels further away from that null. Therefore, although the calibration solutions will generally be improved because of the reduction in degrees of freedom through spectral regularisation, areas with fast spectral beam variation will suffer from gain errors in the channels adjacent to the null. 

The first solution that may come to mind is to remove the calibration beam from the algorithm, as this would remove the spike in the ratio between the true and calibration beam. However, this would mean that any beam variations smaller than a solution interval in time, frequency or space (i.e. a calibration cluster) cannot be modelled. Therefore, a beam model that is correct to the first order is preferred over removing the beam model altogether, although erroneous solutions may need to be excised after calibration. Furthermore, we illustrate the effect here for a large beam error, but the same can occur if the primary beam model is perfect, as long as the null is present. In this case, the spike results from the low signal-to-noise ratio at the null, and the adjacent channels are still perturbed. Excising the parts of the data where these nulls occur is also not a good solution, because it leads to a high loss of data. This is because there can be multiple off-axis sources, so there can be many time and frequency intervals where one of these crosses a null. Furthermore, because this would involve excising entire stations rather than single baselines, many data can be affected.

\subsection{Robust spectral regularisation}
As an alternative to the current form of spectral regularisation, we propose a modified version of the Gaussian-smoothing algorithm that naturally fits in the current approach through reweighting the regulariser. Effectively, this reweighting scheme changes the spectral smoothness assumption; without reweighting, the assumption is that the gains are smooth in frequency.  However, with the reweighted regulariser, we assume that the beam \textit{response} (i.e. the product of the gain and the calibration beam) is a smooth function of frequency. This is a more accurate assumption, because the true beam has shallower nulls than the calibration beam, due to variations within the station that break its symmetry. In our simulations, these variations consist of switched-off elements. In general, intra-station variations occur due to a multitude of additional reasons, such as element-gain variations or variations in mutual coupling between antennas at the centre or the edge of a station \citep{elder_investigating_2024}. 

Concretely, we propose a new set of weights given by the old weights and the calibration model, i.e. 
\begin{equation}
    \mathmat{W}'_{p,k} = \mathmat{W}_{p,k}\odot\sum_{q,l, d\in k}\left|\mathmat{A}_{p,l,d}\mathmat{C}_{p,q,l,d} \mathmat{A}_{q,l,d}^{H}\right|,
    \label{eq:DDECal_weights}
\end{equation}
\noindent for each frequency interval ${\nu_\mathrm{sol}}$. The sum in \Cref{eq:DDECal_weights} acts as a model-flux weighted estimate of the beam response in direction $k$. The weights are high when the beam-attenuated calibration model has a high absolute value, but low when there are nulls in the calibration beam. Therefore, the high-value gain corrections (the spike in \Cref{fig:beam_cartoon}) are down-weighted during convolution, such that they have a lower impact on adjacent channels. This leads to a regularized gain solution that is not as sensitive to large deviations. We illustrate the reweighted gain update with the red line in \Cref{fig:beam_cartoon}. Outside of the channels that fall exactly on the beam null in the calibration beam, the reweighted method (red line) approximates the ideal gain solution (orange line) much better than the unregularised (blue) and standard regularised (green) lines in the top panel. As a result, it outperforms the other methods in approximating the true station response (thick tan-coloured line) in the bottom panel.

We refer to the reweighted method as the `robust' weighting scheme (and compare it to the `standard' weighting scheme, which is not reweighted by the model flux) and explain how it is tested on simulated data in the next sections. We do emphasise that this reweighted calibration scheme is robust against errors in the spectral smoothness assumption, but does not remedy errors on scales smaller than a solution interval or calibration cluster. A DD-gain calibration method still divides the sky in calibration clusters, effectively enforcing a two-dimensional step-function on the sky, and similarly divides the time axis in discrete solution intervals. The robust calibration scheme is therefore mainly effective against beam modelling errors that change rapidly as a function of frequency.

\section{Simulations} \label{sec:sims}
Because beam changes resulting from gain-deviations on the individual antenna element level have not previously been investigated for full-scale LOFAR-HBA observations, we have implemented the simulation and calibration pipeline in a Python code named \codename\ (Station Heterogeneity Impact on Multi-dimensional beam-Modelling Errors simulatoR and calibratoR)\footnote{\href{https://github.com/Stefanie-B/shimmerr}{https://github.com/Stefanie-B/shimmerr}}. This codebase is able to simulate any hierarchical interferometer, as long as the European Terrestrial Reference System coordinates of the individual antenna elements are known. It has been highly parallelised to allow for simulations of full LOFAR datasets and can simulate beam errors stemming from broken or miscalibrated (sets of) antenna elements, and pointing errors. When these effects are disabled, the elements at each level of the beam-forming hierarchy become identical, such that their beams only need to be computed once. The computation can then be sped-up by calculating a single beam and simulating the full beam by applying the array factor (\Cref{eq:array_factor}, where $m$ can iterate over antennas or tiles, based on the level in the hierarchy). 

\begin{figure*}
    \centering
    \includegraphics[width=0.9\linewidth]{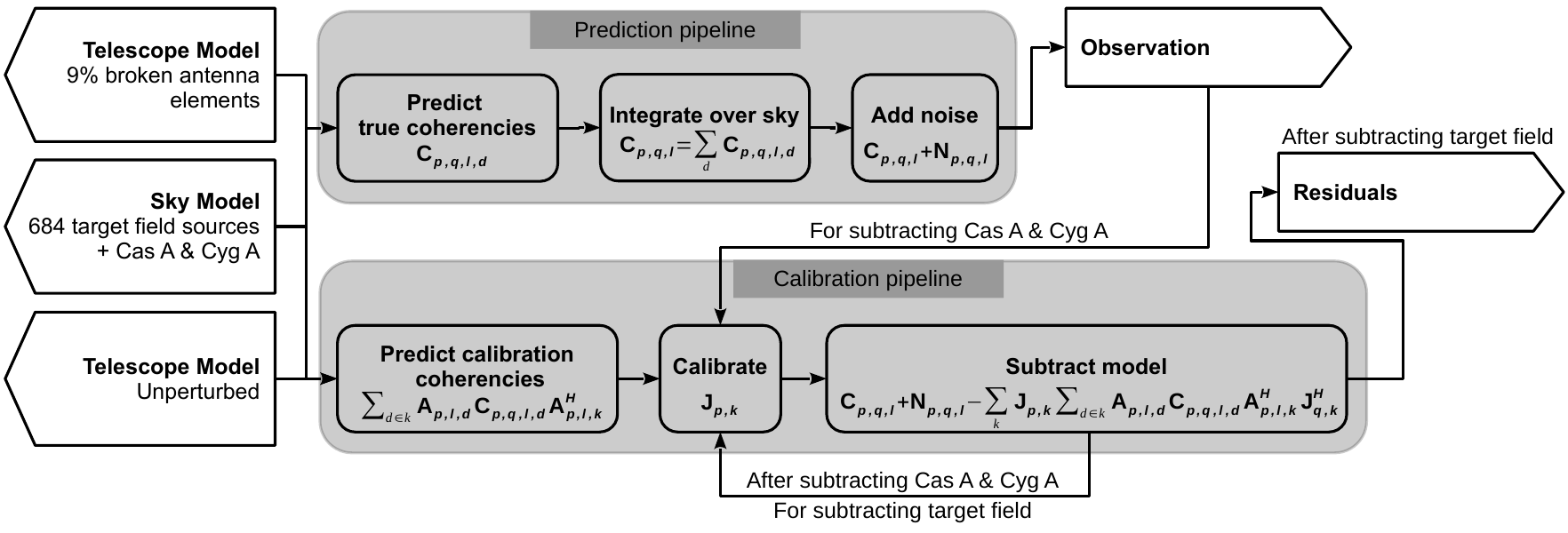}
    \caption{Pipelines used for forward simulation of the visibilities and the calibration. Boxes with an arrow on the left are inputs, boxes with an arrow on the right are outputs and rounded boxes are processing steps.}
    \label{fig:sims_pipeline}
\end{figure*}

To reduce the computational complexity of the forward prediction and because the behaviour of nulls in the beam is not expected to behave qualitatively differently between polarisations, we have chosen to implement \codename\ for only a single polarisation, which we treat as Stokes I, currently. The pipelines for visibility prediction and calibration are illustrated in \Cref{fig:sims_pipeline}. In the remainder of this section, we describe how the different parts of the visibility prediction and calibration simulations are implemented and the settings we have chosen for the simulated observations and calibration.

\subsection{Telescope Model}\label{subsec:sims_telescope}
\begin{figure*}
    \centering
    \includegraphics[width=\linewidth]{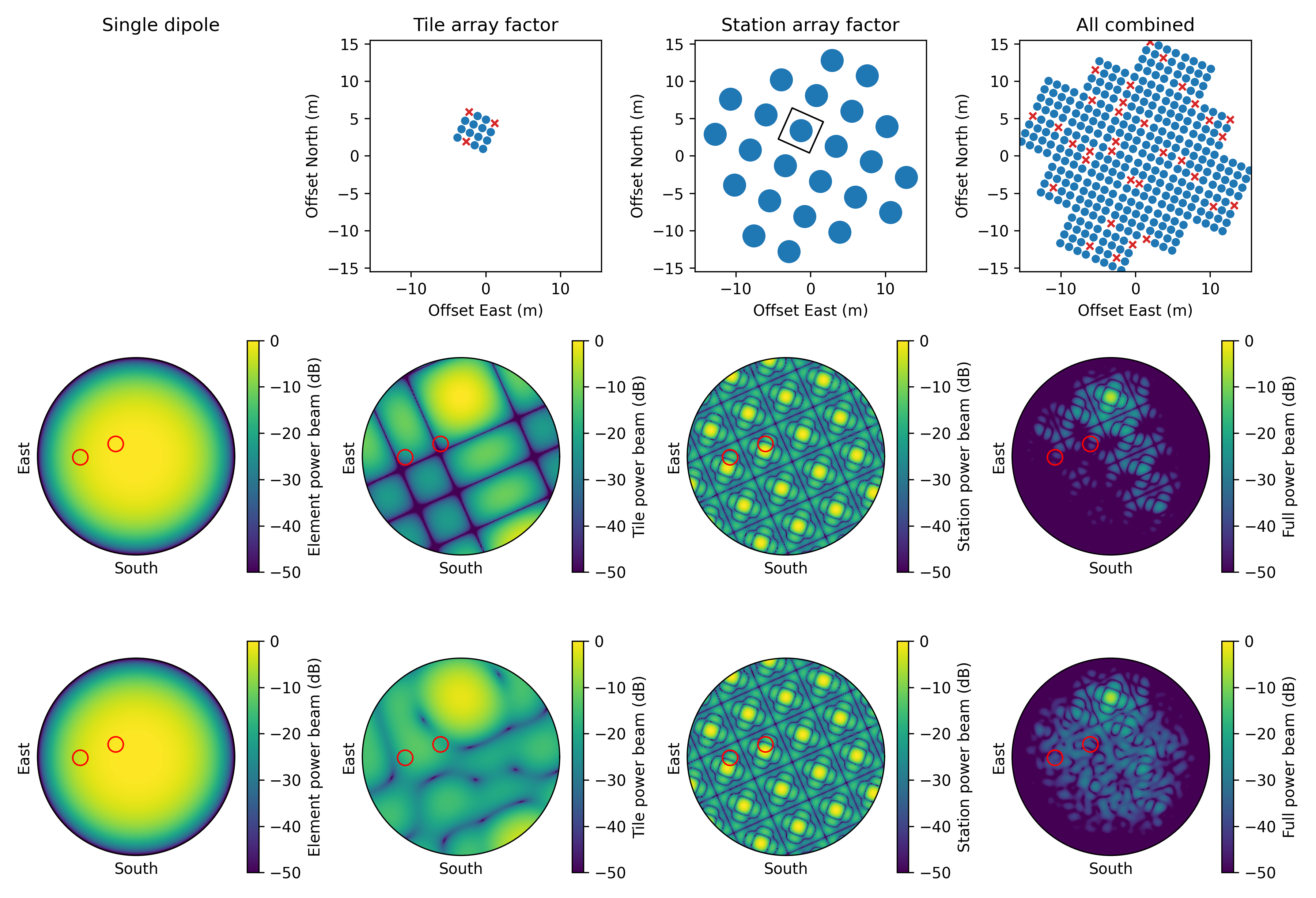}
    \caption{Illustration of the hierarchy of the LOFAR-HBA beam for station CS001HBA0 pointed to the NCP at 03:00:00.00 on 28-06-2014 UTC at 140~MHz. Top row: positions of the elements, where small circles and red crosses indicate antenna elements and big circles indicate tiles, respectively. Middle row: Calibration beams, where no elements are broken. Bottom row: True beams, where the antennas marked by red crosses are missing. From left to right: the element beam, the per-tile array factor, the station array factor if tiles are used as elements, and the full station beam. The full station beam is the beam that is obtained if all three beams in the columns to the left are taken into account. The black square in the top plot of column three shows the tile that is plotted in column two. The red circles in the bottom two rows show where the two bright sources Cassiopeia A and Cygnus A are.}
    \label{fig:beam_hierarchy}
\end{figure*}

The goal of the forward simulations is to generate a set of `true' visibilities that describe the behaviour of the telescope if there are station-beam perturbations. The basic building block of a LOFAR-HBA station is a cross-bowtie antenna. Sixteen of such antennas form a `tile' (see the first panel in the top row of \Cref{fig:beam_hierarchy}). Tiles are analogue beamformed \citep{van_haarlem_lofar_2013}. A group of 24, 48, or 96 tiles -- depending on the type of station -- are digitally beamformed to create a station beam. The digitally beamformed station responses to the sky are correlated to create visibilities, as described in \Cref{eq:fullsky_RIME}.

An example of the beams for different beam-forming hierarchy levels within the station are shown in the middle row of \Cref{fig:beam_hierarchy}. In the first column, we show a beam model for the antenna element, based on the beams presented in Chapter 2 of \citet{heald_low_2018}. This beam does not need to be very accurate, because its large width in comparison to the final station-beam width results in limited impact on calibration. The second column shows the array factor of a tile (not multiplied by the element beam). This beam can be steered through analogue beamforming. In this simulation setup, it is steered towards the North Celestial Pole (NCP) field. The third column shows the array factor of the station (with tiles as elements), not multiplied by the tile beams. The final column shows the full station beam. If no perturbations to the individual elements are present, this is equal to the product of the three beams to the left of it. LOFAR-HBA stations are rotated with respect to one another to mitigate the effect of strong sidelobes and nulls \citep{wijnholds_calibratability_2011}, resulting in differing beam patterns between stations.

\subsection{Beam Perturbations}\label{subsec:sims_brokenelements}
Diagnostic information on the performance of elements is only available at the tile level in LOFAR-HBA observations, because of the analogue beamformer. Tiles that have degraded past a certain point will be excised upon detection, which means that they are not included in the digital beamformer and not included in the simulated calibration beam. However, if an antenna within a tile breaks, this can only be detected as a different tile response. While the tile responses of LOFAR have been calibrated during commissioning, such calibration rounds can be out of date. Currently, typically up to three antenna elements are broken or malfunctioning per tile (Menno Norden, private communications). 

\begin{figure}
    \centering
    \includegraphics[width=\linewidth]{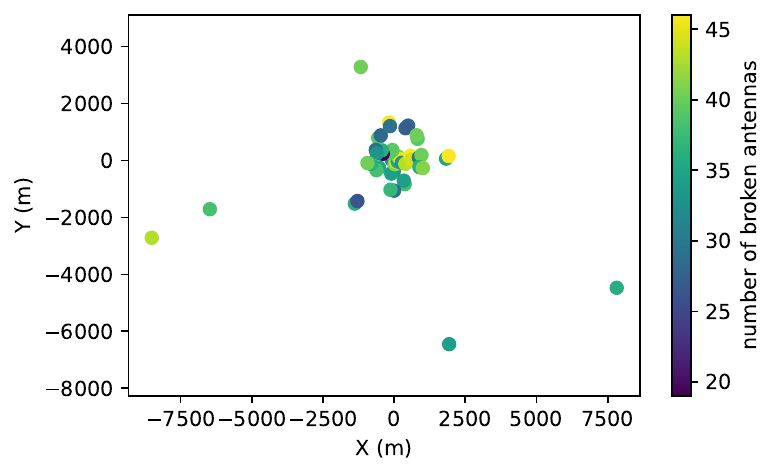}
    \caption{Layout of the stations included in the simulation. The y-axis shows the North-South offset from the centre of the interferometer, while the x-axis shows the East-West offset. The stations are colour-coded by the number of broken antenna elements in the true beam. The total number of antennas in a station is 384.}
    \label{fig:array_map}
\end{figure}

The exact distribution of broken elements is unknown. Therefore, we simulate beam errors by randomly switching off a number of elements, by setting their element gains in \Cref{eq:DDECal_A} to zero. Throughout our simulations, $N\sim\mathcal{U}(0,3)$ elements are switched off in each tile, where $\mathcal{U}(0,3)$ is the discrete uniform distribution between 0 and 3. This results in an average of 1.5 antennas, or 9 per cent of the tiles being simulated as `broken'. We randomise the broken antennas for each tile in each station, such that different stations have a different configuration of broken elements. Because the number of broken antennas per tile is randomised, different stations also have varying numbers of broken elements, as shown in \Cref{fig:array_map}.

The effects on the beam, resulting from a random realisation of removing antennas, is illustrated in the last row of \Cref{fig:beam_hierarchy}. As is clear from the second column, switching off three antenna elements drastically alters the shape of the tile beam, removing and altering nulls. Because the tile beams are no longer identical, the nulls and sidelobes of the full station primary beam also change. This has a limited effect near the phase centre (bright yellow spot), but does change when off-axis sources pass through nulls. We illustrate where such off-axis sources can be compared to the phase centre through red circles at the positions of Cas A and Cyg A.

It is important to emphasise that, although we focus on broken elements in this work, there are other effects that break the regular structure of a phased array station and create similar distorted beam patterns to those shown here. For example, mutual coupling \citep{elder_investigating_2024,ohara_uncovering_2024} will affect the beam by changing the patterns of individual antenna elements. Performance (e.g. gain) differences between functioning elements \citep{chokshi_necessity_2024} will have a similar effect to setting the gains of several elements to zero. These errors can also change over time, for example, by the low-noise amplifiers being affected by on-site temperature and humidity variations. Additionally, such variations can occur at the tile (analogue-beamformed) beam level rather than the individual antenna element level. However, because we expect broken elements to have a dominant effect, we demonstrate the resulting calibration problems arising from such beam errors.

\subsection{Sky Model}\label{subsec:sky_model}
For sources in the main lobe, we use a model of the North Celestial Pole (NCP) field, which is the most studied field in the LOFAR-EoR key science project \citep{patil_upper_2017,mertens_improved_2020,mertens_deeper_2025}. The model is a $10^\circ\times 10^\circ$ field of view containing 684 flat-spectrum point sources, that dominate the power spectrum of the field, distributed over 24 calibration clusters (see \citealt{brackenhoff_ionospheric_2024}).

For the off-axis sources, we use two bright radio sources: Cassiopeia A (Cas A) and Cygnus A (Cyg A). These sources are approximately 30$^\circ$ and 50$^\circ$ away from the target direction, respectively. These are important to include, because their residuals after DD-corrected model subtraction currently dominate the 21-cm signal power spectra of the NCP field \citep{gan_statistical_2022,munshi_first_2024,ceccotti_spectral_2025}. We base our models of Cas A and Cyg A on those provided by ASTRON\footnote{\href{https://github.com/lofar-astron/prefactor/blob/master/skymodels/A-Team_lowres.skymodel}{https://github.com/lofar-astron/prefactor/blob/master/skymodels/A-Team\_lowres.skymodel}}, but replace Gaussian components by point sources with the same total brightness. {\scriptsize DDECAL} calculates the model for a Gaussian source by simulating it at its centroid and scaling its brightness according to baseline length. This means that DD (e.g. beam) effects across an extended source are not taken into account. Therefore, we would not obtain additional DD information by modelling them as Gaussians.

Two variations of the model are considered: a single point source for the total brightness of the A-team source (the `single point-source model') and a version where Cas A and Cyg A consist of 9 and 5 point sources respectively (the `multi-point source model'). Comparing the results obtained between these models allows for comparing the extreme case of a single point-source shifting in and out of a null to the more general case of an extended source (represented by multiple point-sources), without needing to simulate Gaussians. 

\subsection{Prediction Pipeline}
\Cref{eq:DDECal_A} is used to forward simulate the beam. Since the chromaticity of the element beam is limited, we model it as an achromatic beam $U(\mathvec{\hat{s}})$ that is identical for all antennas. The station beam can be computed with and without removing elements. Setting all element gains to unity results in the calibration beam, whereas setting some element gains to zero results in the perturbed `true' beam. 

We use the true beam (with switched-off elements) as the propagation matrix in \Cref{eq:fullsky_RIME} to simulate visibilities. The thermal noise is created by independently drawing the real and imaginary part of the noise from a Gaussian with standard deviation $\sigma=\text{SEFD}/\sqrt{2\Delta\nu\Delta t}$ [Jy]\footnote{The factor two is introduced by independently drawing the real and complex part of the noise.}, where $\Delta\nu$ and $\Delta t$ are the spectral and temporal resolutions of the simulation in Hertz and seconds, respectively. SEFD is the system equivalent flux density in Jansky. The visibilities created in this way (with the true beam and noise) are treated as the observation's raw uncalibrated data.

For the calibration model, we subsequently use the model beam (that does not contain broken elements). We sum the coherencies per calibration patch to create model visibilities for each direction, and do not add thermal noise.

\subsection{Simulated observations}
\Cref{tab:prediction_settings} lists the most important settings used for the simulated observations. These settings have been chosen to mimic typical LOFAR-EoR observations. Each observation consists of a 12-hr rotation synthesis and has the same time resolution as as is used in LOFAR-EoR DD-gain calibration. The total bandwidth matches a LOFAR-EoR redshift bin and the spectral resolution is one channel per sub-band, equal to the calibration resolution on real data. The SEFD has also been chosen to match the noise level of these observations \citep{mertens_improved_2020}. Finally, the telescope model has been configured to match the configuration of a typical LOFAR-EoR dataset. The core stations (CS), which consist of pairs of stations of 24 tiles, are configured to act independently. This means that one CS acts as two independent but closely spaced stations, labelled by a 0 and 1, respectively (for example: CS002HBA consists of CS002HBA0 and CS002HBA1). The remote stations (RS), which are usually comprised of 48 tiles, have their outer 24 tiles switched off to match the shape of a CS. This makes the core and remote stations identical in our simulations\footnote{In real observations, the outer tiles of the RS receive a zero-weight such that they do not directly contribute to the observation. However, because they are still present, they do electrically couple to the inner tiles, distorting the beams in a different way than the CS.}.

\begin{table}
    \centering
    \caption{Prediction settings for the simulated observations.}
    \label{tab:prediction_settings}
    \begin{tabular}{l c}
        \hline
        Parameter & Setting\\
        \hline
        \multirow{2}{*}{Telescope} & LOFAR HBA -- CS (dual mode) \\
        &+ RS (in CS configuration)\\
        Missing elements & 9\% of station\\
    Pointing & RA = $0^\mathrm{h}00^\mathrm{m}00^\mathrm{s}$, Dec = $+90^{\circ}00'00''$\\
        \multirow{3}{*}{Start times (UTC)}&27-12-2013 18:00:00 (N1)\\
        &27-03-2014 18:00:00 (N2)\\
        &27-06-2014 18:00:00 (N3)\\
        \multirow{3}{*}{Sky model} & 684 point sources in the central $10\times 10^\circ$\\
        & Cassiopeia A (1 or 9 point sources)\\
        & Cygnus A (1 or 5 point sources)\\
        Bandwidth & 135-147~MHz\\
        Frequency resolution & 195~kHz\\
        Time resolution & 10~s\\
        Total duration & 12~hr\\
        SEFD & 4.2 kJy\\
        \hline
    \end{tabular}
\end{table}

\subsection{Calibration}
A telescope without missing elements is used to create model coherencies for calibration purposes. Because we have performed the forward prediction with unity gains for all except the switched-off elements, and because we do not assume band-pass errors, we only consider a DD-gain calibration step for source subtraction. If both direction-independent (DI) and DD-gain calibration are performed, errors in the DI gains could be absorbed in the DD-gains, such that the results become less interpretable. Calibration is performed in two steps, because the station beams change faster as a function of time and frequency for sources far from the target direction. Because of this, Cas A and Cyg A should be calibrated for and subtracted with a smaller time interval per solution than used for the target field\footnote{In the LOFAR NCP field, these two steps are performed simultaneously with {\scriptsize SAGECAL-CO}. However, setting two different solution intervals is not possible with the {\tt directionsolve} method of {\scriptsize DDECAL}.} \citep{gan_assessing_2023,munshi_first_2024}. The first round of calibration is done at time interval of 2.5 minutes per solution and in three directions: Cas A, Cyg A, and the target field for which we assume a single direction. Cas A and Cyg A are removed with the gain-calibration solutions for their respective directions applied to their respective sky-models. The residuals from this calibration step are used as an input for the second round of calibration using a solution interval of 20 minutes, in which the target field is divided into 24 gain-calibration directions. In the LOFAR-EoR key science project, DD-gain calibration is performed on a set of longer baselines ($250-5000\lambda$), to avoid absorption of the large-scale 21-cm signal in the calibration solutions \citep{patil_systematic_2016,mevius_numerical_2021}. The analysis is done on shorter baselines ($50-250\lambda$). Therefore, we also calibrate on these longer baselines only, but analyse all baselines between $50~\lambda$ and $5000~\lambda$. We omit three stations, that are a part of fewer than three baselines in this range. An overview of the calibration settings is listed in \Cref{tab:calibration_settings}.

\begin{table}
    \centering
    \caption{Calibration settings for subtracting the sky model}
    \label{tab:calibration_settings}
    \begin{tabular}{l c c}
         \hline
         \multirow{2}{*}{Parameter} & Value for & Value for\\
         &Cas A and Cyg A & target field \\
        \hline
         Number of calibration directions $N_k$ & 3 & 24\\\
         Solution time interval (min) & 2.5 & 20\\
         Baseline lengths calibration ($\lambda$) & \multicolumn{2}{c}{250--5000} \\
         Solution frequency interval (kHz) & \multicolumn{2}{c}{195}\\
         Number of iterations & \multicolumn{2}{c}{50}\\
         Reference station & \multicolumn{2}{c}{CS002HBA0}\\
         \hline
    \end{tabular}
\end{table}
\section{Results} \label{sec:results}
In this section, we analyse the effect of the beam errors on the calibration results. First, the expected and computed gains are discussed to build more intuition on the spectrotemporal signatures of the beam errors. With this in mind, the residuals after subtracting all sources with the computed DD-gains are inspected for various spectral smoothness kernel widths for both the standard and the robust regularisation scheme. Finally, the effect of the errors in image space is discussed.

\subsection{Gain Solutions} \label{subsec:results_gains}
Because the Earth rotates during an observation, the spatial beam pattern shown in the last column of \Cref{fig:beam_hierarchy} moves across the sky and changes shape (or, in the case of the NCP, rotates across the sky). Sources that are not in the station-beam main lobe pass through multiple nulls and sidelobes. Additionally, the position and width of the station-beam sidelobes depend on the spacing between antennas in units of wavelength. This causes the beam lobes to be narrower at higher frequencies, and wider at lower frequencies. These two effects combined define the beam as a four-dimensional structure with two spatial, one spectral and one temporal dimension. For a given point in the sky, this reduces to a two-dimensional dynamic spectrum for the gain $|\mathmat{J}_{p,k}(\nu,t)|$.

\begin{figure}
    \centering
    \includegraphics[width=\linewidth]{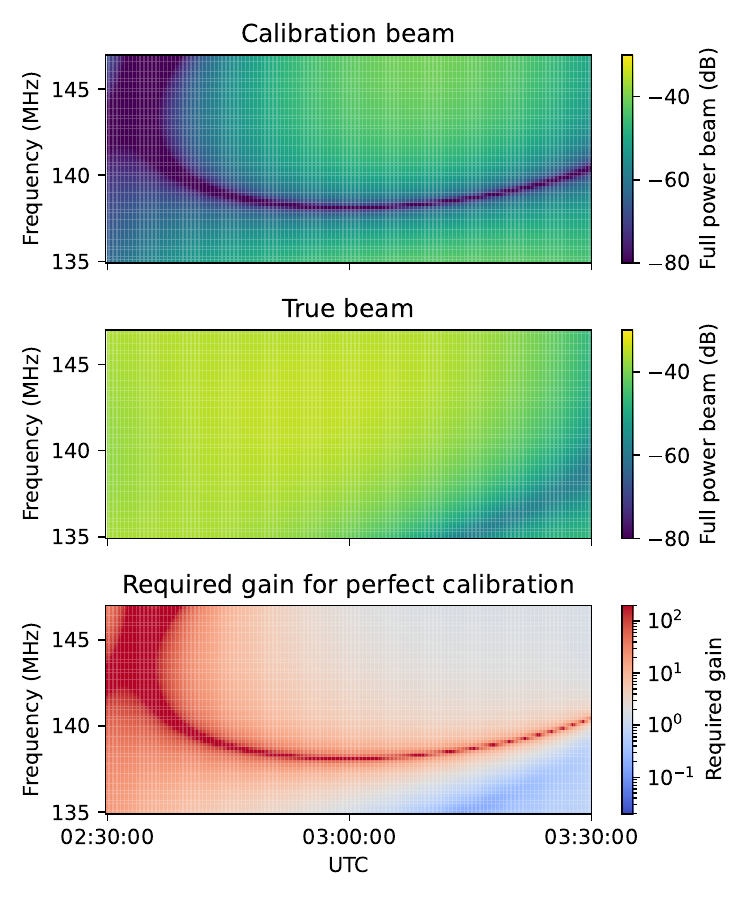}
    \caption{Effect of beam errors in frequency and time for station CS001HBA0 on 28-06-2014 UTC (N3) in the direction of Cas A. This is the same station as in \Cref{fig:beam_hierarchy}. Top: Expected beam response (calibration beam), middle: Beam response with 32 broken elements (`true' beam), bottom: gain required to fully subtract Cas A at the same resolution as the data.}
    \label{fig:res_beam_vt}
\end{figure}

\Cref{fig:res_beam_vt} shows the dynamic spectrum of the station power beam\footnote{As is customary in literature about beams, we use a `power beam' in the plots. The power beam is $|\mathmat{A}_{p,k}(\nu,t)|^2$.} from \Cref{fig:beam_hierarchy} for the direction of Cas A. The top panel of \Cref{fig:res_beam_vt} shows the calibration beam (without switched-off elements), and the middle panel shows the true beam (with switched-off elements). When calibrating, the goal is to approximate the true beam with the product of the gain and calibration beam. The gain that would perfectly do this for a single point-source is equal to the ratio between the true beam and the calibration beam. This ideal gain is shown in the bottom panel of the figure.

Gain-calibration algorithms do not reach this ideal gain for two reasons: the finite calibration resolution needed to reach a minimum signal-to-noise ratio for the gain solutions (gain-discretisation), and the non-convex nature of the optimisation problem leading to solution errors. Gain-discretisation introduces errors because variations on scales smaller than a solution interval cannot be captured. In other words, a single matrix $\mathmat{J}_{p,k}$ cannot fully compensate for modelling errors in $\mathmat{A}_{p,l,d}$ as a function of $l$ (time and frequency indices within a solution interval) or $d$ (source direction within a calibration cluster). Therefore, if the variation of the station-beam in space, time, or frequency is not well-modelled within a solution interval, these intra-interval errors remain present after calibration. Additionally, the ideal gain may have spectral variations on scales that are suppressed by the low-pass filter. This effect is especially strong near nulls in the calibration beam, where the desired gain peak approaches infinity in case the true beam is non-zero. Around a null, the ideal gain solution has a very rapid rise and fall. However, such changes cannot be captured with a smooth gain solution. This rise and fall is visible as the horizontal red streak in the bottom panel of \Cref{fig:res_beam_vt}. In addition to these resolution-based effects, there is a risk of non-convergence (after the maximum number of iterations) or convergence to a local minimum in the optimisation problem. This occurs because the final gain solution $\mathmat{J}_{p,k}$ depends on the intermediate estimates of the gains of all stations and all directions. Therefore, an error in one direction for one station can potentially affect the estimates of all other gains. 

\begin{figure}
    \centering
    \includegraphics[width=\linewidth]{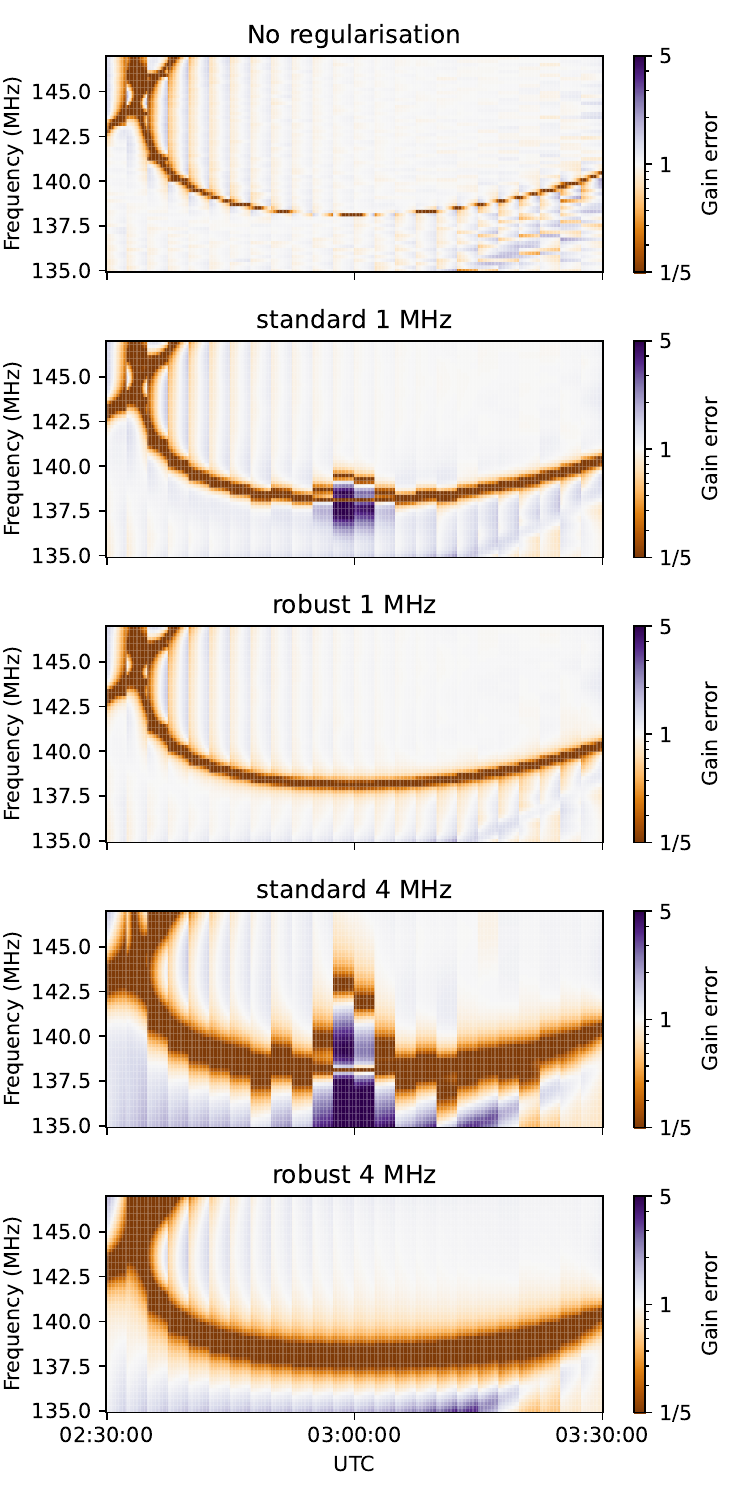}
    \caption{Ratio between the gain computed using {\scriptsize DDECAL} and the required gain in the direction of Cas A obtained with the point-source model for Cas A and Cyg A. Top to bottom: Result without regularisation, with a 1~MHz kernel and the standard regularisation method, with a 1~MHz kernel and the robust regularisation method, with a 4~MHz kernel and the standard regularisation method, and with the 4~MHz kernel and the robust regularisation method. White indicates an accurate gain solution, purple an overestimation of the gain and orange an underestimation. The station and time match \Cref{fig:res_beam_vt}.}
    \label{fig:res_gain-comparison}
\end{figure}

With a noisy 12~hr simulation in hand, we run {\scriptsize DDECAL}. The errors in the gain-calibration solutions in the direction of Cas A are shown in \Cref{fig:res_gain-comparison}. In all panels, the single point-source models are used for Cas A and Cyg A, but the spectral regularisation settings vary between panels. The gain error is computed by dividing the amplitude of the calculated gain by the amplitude of the ideal gain (i.e. the ratio of the true and calibration beams shown in the bottom panel of \Cref{fig:res_beam_vt}). The top panel of \Cref{fig:res_gain-comparison} shows the result if no spectral regularisation is used during calibration, and generally follows the expected gain behaviour shown in the bottom panel of \Cref{fig:res_beam_vt}. The unregularised solution is not useful in practice due to overfitting of the data, as is visible as the rapid variations in the lower right corner of the plot. We include the unregularised solution here purely as a reference plot, to illustrate the effect that regularisation has. The discretisation of the gains, discussed above, is clearly visible as steps in the gains every 2.5 minutes. Additionally, the gains are underestimated along the null in the calibration beam, as is evident from the orange regions in the same areas where the null was identified in \Cref{fig:res_beam_vt}. Because the ideal gain approaches infinity here, this underestimation is expected. However, because the null in the calibration beam is narrow, this affects few channels, that could be flagged later.

The second and third rows of \Cref{fig:res_gain-comparison} both use a 1~MHz wide smoothness scale. However, the second row uses the standard regularisation method and the third the robust one. The fourth and fifth rows show the same, but for a wider 4~MHz spectral smoothing kernel. The standard smoothing method (the second and fourth row) shows a gain-solution artefact at 03:00:00 that is not present for the robust regularisation method. The gains near the null are overestimated, with an underestimation exactly on the null. Additionally, there is an underestimation of the gain near the null that is shifted towards the centre of the band (at $\sim$140~MHz and $\sim$143~MHz for rows two and four respectively). The overestimation at the null surrounding an underestimation is expected from applying a low-pass filter to a peak in the gain solutions, as illustrated in \Cref{fig:beam_cartoon}. The location of the gain-calibration error is also in line with expectations, as the null near 03:00:00 changes more rapidly in the spectral direction than the temporal direction. We attribute the underestimation towards the centre of the band to a local minimum. The low pass filter increases the gains here, due to the overestimated gains near the null. This can affect the gains further away in frequency as well. Therefore, in the next iteration, a least-squares solution is found where the gains towards the centre of the band (at $\sim$140~MHz and $\sim143~$MHz for rows two and four respectively) are lowered, to `insulate' the rest of the band from the effect of the badly calibrated null during smoothing. The same does not happen on the other side of the null, away from the centre, because there are fewer channels on this side of the null. Therefore, there are not enough channels that would be affected to create the `insulation' effect.

However, when robust regularisation is applied (rows three and five of \Cref{fig:res_gain-comparison}), the artefact disappears and the behaviour of the null at 03:00:00 is very similar to the rest of the observation. In this part of the gain dynamic spectrum, robust regularisation clearly shows fewer errors than the standard method and limits the number of channels affected by the effect of the null in the calibration beam, as intended. There are still clear errors in the recovered gains, however. Furthermore, the wider kernel seems to both increase the number of channels with errors (wider orange streaks) and the amplitude of errors that were already present (stronger purple and orange shading near 135~MHz), increasing the bias. 

\subsection{Residual visibilities}
\begin{figure*}
    \centering
    \includegraphics[width=\linewidth]{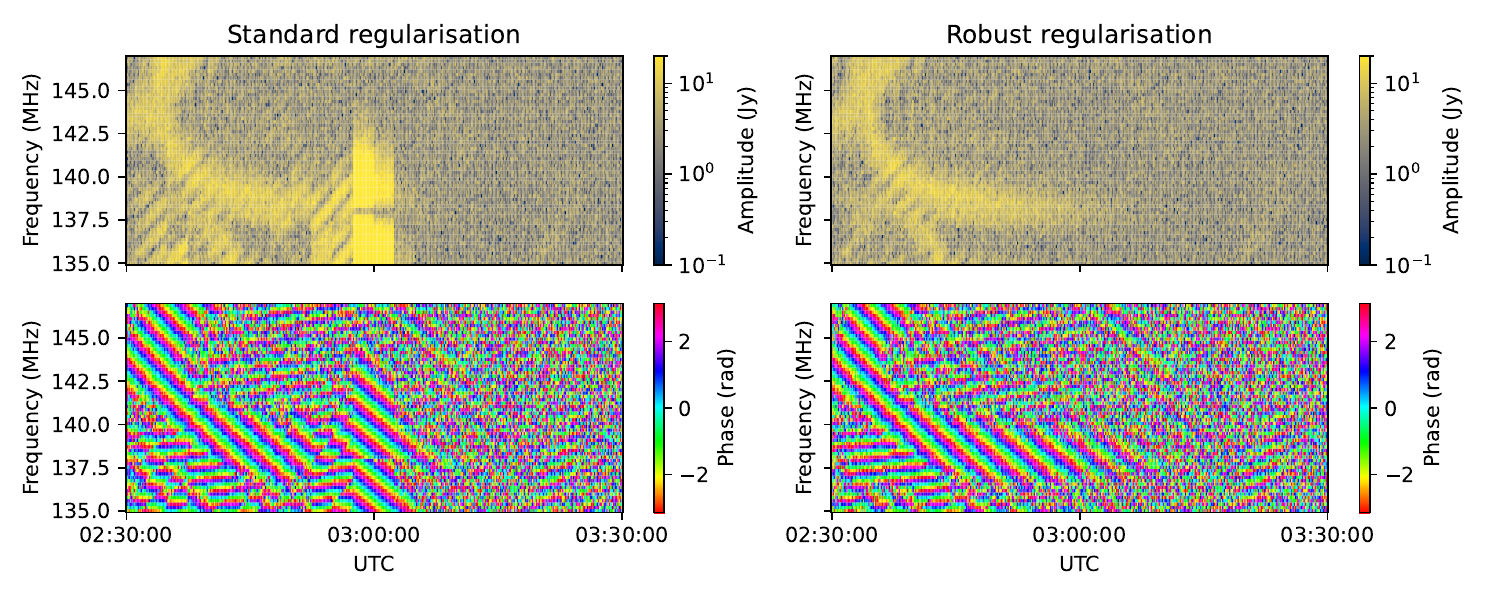}
    \caption{Residual visibilities for the baseline between CS001HBA0 and CS003HBA1 on 28-06-2014 (N3) after all sources have been subtracted with solutions created with the 4~MHz kernel. The times and frequencies match \Cref{fig:res_beam_vt,fig:res_gain-comparison}. Left: Standard {\scriptsize DDECAL} algorithm, right: the robust method. Top: amplitudes, bottom: phases. Ideally, these residuals should be noise-like.}
    \label{fig:res_single_bl}
\end{figure*}

\begin{figure*}
    \centering
    \includegraphics[width=\linewidth]{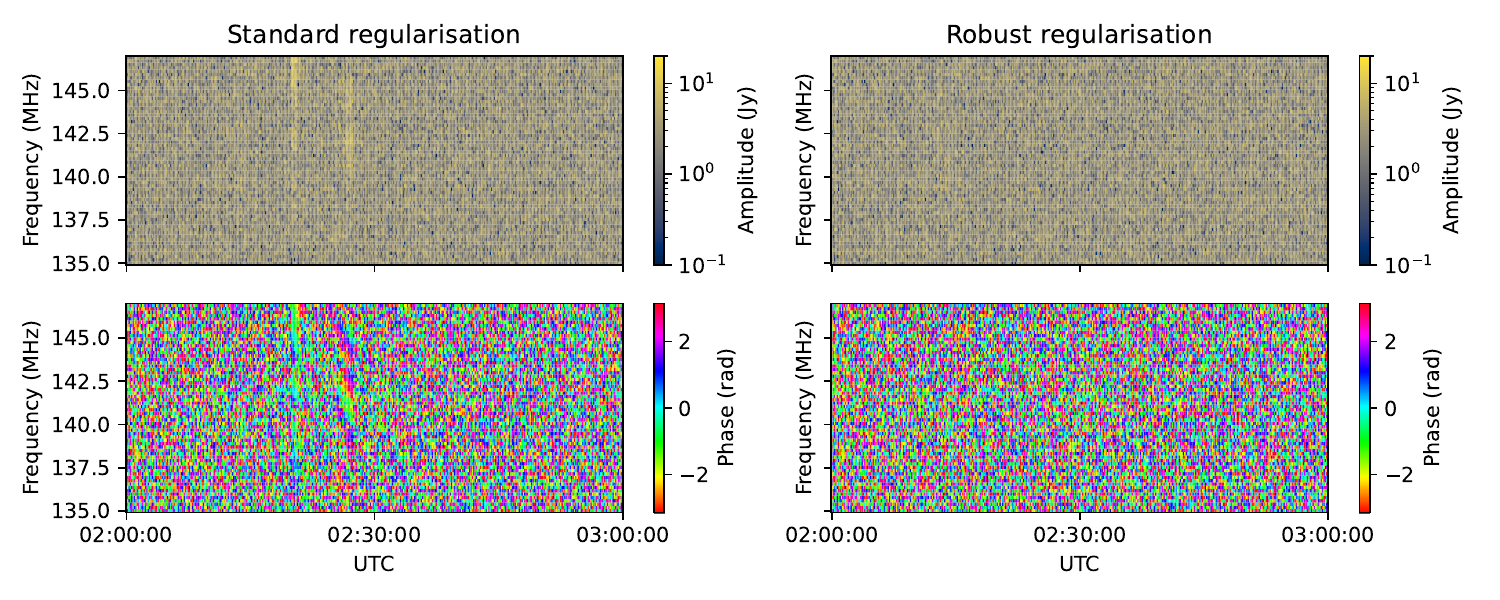}
    \caption{Similar to \Cref{fig:res_single_bl}, but for the baseline between CS002HBA1 and CS007HBA1 on 28-12-2013 (N1). Here, the robust regulariser is able to subtract the sky sources down to the noise level.}
    \label{fig:res_single_bl2}
\end{figure*}

We now turn our attention to the residual visibilities that are found after subtracting all sources with their respective DD-calibrated gains. Because there are no differences between the simulated and calibration sky models, a perfect calibration would result in the residual visibilities being completely noise-like\footnote{Due to this being a non-linear model, a small bias is expected, but this is negligible compared to the error level.}. The accuracy of the calibration can therefore be tested by comparing the residual visibilities to noise. In \Cref{fig:res_single_bl,fig:res_single_bl2}, the dynamic spectra of the residual visibilities on single baselines are shown. One of the stations that comprise the baseline in \Cref{fig:res_single_bl} is the same station as used for \Cref{fig:beam_hierarchy,fig:res_beam_vt,fig:res_gain-comparison}, such that the nulls seen in \Cref{fig:res_beam_vt,fig:res_gain-comparison} can be associated with effects seen in \Cref{fig:res_single_bl}. The left-hand columns of \Cref{fig:res_single_bl,fig:res_single_bl2} show the residual visibilities after subtracting the gain-calibrated sky model with a 4~MHz kernel and the standard method, and the right-hand columns show the same, but with the robust gain-calibration method. The top panels show visibility amplitudes, and the bottom panels show visibility phases. 

The gain errors seen in \Cref{fig:res_gain-comparison} clearly have a detrimental effect on the subtraction of Cas A, because the areas that have large gain errors in \Cref{fig:res_gain-comparison} also exhibit an increased amplitude and structure in the phases in \Cref{fig:res_single_bl}. There are also other areas with large residual visibilities, such as the streak below 140~MHz between 02:35 and 02:45. This artefact coincides with a null in the direction of Cyg~A, the other bright off-axis source. Similar to what was found for the gains, robust gain-regularisation yields an improvement in the calibration results. It reduces the artefact at 03:00:00, both in terms of amplitude and number of frequency channels affected by it. However, even the robust regularisation method does not completely remove Cas A and Cyg A. These residuals are strongly time- and baseline-dependent, however. This can be seen in \Cref{fig:res_single_bl2}, where robust regularisation is able to subtract all sources down to the noise level.


The main improvement between the standard and the robust method is that the standard method can introduce large outliers in the gains, which can in turn result in large outliers in the visibility residuals. However, typically, such outliers in non-transient visibility datasets could be remedied with outlier-based data excision. Therefore, a more realistic comparison would be to compare the two methods after excision of the outliers. A challenge is that our simulations are solely corrupted by beam errors, whereas a real observation might have multiple different sources of outliers in the residual visibilities. Therefore, low-level artefacts specifically introduced by errors in the gain would be more difficult to identify and excise in real data. Nonetheless, we can draw conclusions about the magnitude and number of outliers compared to the thermal noise level, as well as the residual power left after excision using these simulations. Because the outliers introduced by beam errors tend to be correlated both in time and frequency, they can be detected with traditional outlier detection methods such as the scale-invariant rank filter that flags data based on the number of outliers within a slice of data ({\scriptsize AOFLAGGER}, \citealt{offringa_morphological_2012}). A single round of {\scriptsize AOFLAGGER} is, therefore, performed on all residual visibilities.

\begin{figure}
    \centering
    \includegraphics[width=\linewidth]{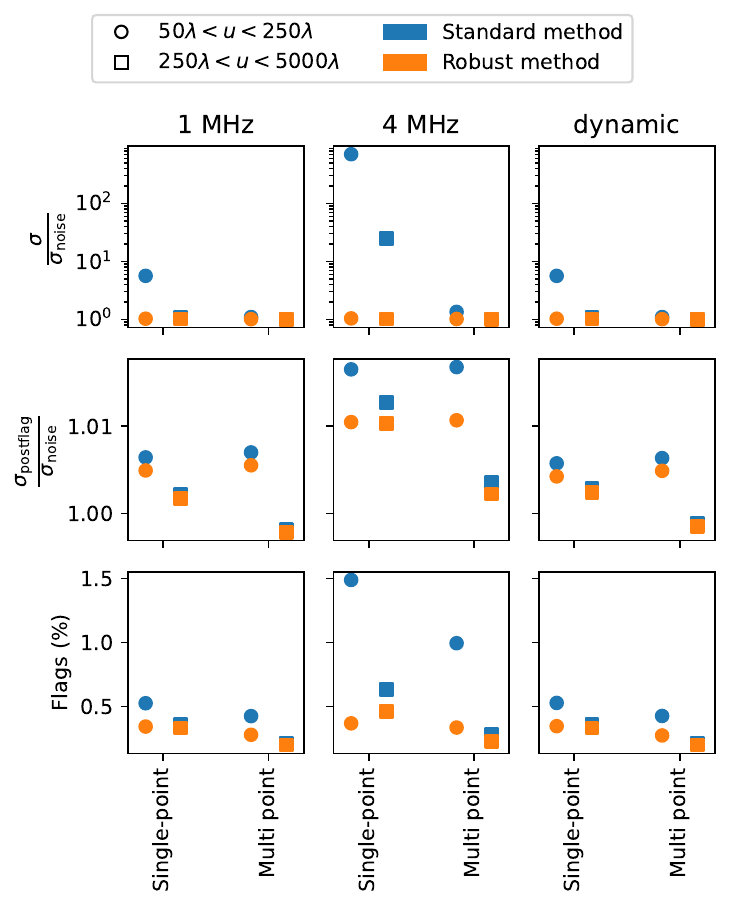}
    \caption{Effect of outlier excision on the dataset, for both the single-point and multi-point off-axis models for the simulation starting on 27-12-2013. Top: standard deviation of the residual visibilities compared to that of the thermal noise, middle: the same after outlier flagging, bottom: percentage of the data that were removed. From left to right, the residuals after calibration with a 1~MHz kernel, 4~MHz kernel and a dynamic kernel with mixed regularisation (1~MHz for Cas A and Cyg A, and 4~MHz for the target field). Blue markers indicate the standard method and orange markers the proposed method. The EoR analysis data, which are baselines $< 250\lambda$ are indicated by circles and the calibration baselines ($> 250\lambda$) by squares.}
    \label{fig:res_flagging_stats}
\end{figure}

Because perfect calibration would result in only Gaussian noise, the effectiveness of the calibration can be judged by deviations from the thermal noise level. Therefore, we compare the standard deviations of the visibilities of a complete 12~hr simulation before and after outlier excision to the thermal noise level of $ \sigma_n=SEFD/\sqrt{\Delta\nu\Delta t}=3.01$~Jy. The standard deviations of the simulations of N1 (that starts on 27-12-2013, see \Cref{tab:prediction_settings}) with the single-point and multi-point models for Cas A and Cyg A are shown in \Cref{fig:res_flagging_stats}. The figure shows the ratio between the standard deviations and noise before and after outlier flagging in the top and middle row respectively. The bottom row shows the percentage of data that was excised by {\scriptsize AOFLAGGER}. We consider the latter statistic, because using fewer flags leads to a lower loss of data, and therefore, a higher sensitivity. This loss of data can be significant, e.g. in LOFAR-EoR science, through various methods of flagging (see \citealt{mertens_deeper_2025}), up to 19\% of data are flagged to excise Cas A and Cyg A (Florent Mertens, private communications). Therefore, attaining the same standard deviation with fewer flags is beneficial.

For all smoothing kernel settings and sky-models, the robust regularisation method matches or outperforms the standard method, both in terms of standard deviations before and after flagging, and in the number of flags that are needed to attain this. Generally, the pre-flagging standard deviation of the standard method is much larger than that of the robust method, such that flagging the data has a much larger effect. 

The different panel columns show various kernel sizes for the calibration. The first and second columns show the 1~MHz and 4~MHz kernels, respectively. Similar to what was found in \Cref{subsec:results_gains}, the 4~MHz kernel performs significantly worse than the 1~MHz kernel. This is somewhat surprising, since \citet{gan_assessing_2023} showed that the 4~MHz kernel performed better on LOFAR-EoR data. We suspect that there is a difference in which kernel width performs better on the main lobe versus off-axis sources. Because the target field does not suffer from beam modelling errors to the same degree as the off-axis sources, the 4~MHz kernel may outperform the 1~MHz kernel for the target field sources, while this is the other way around for the far-field sources. A similar spectral smoothness difference between the main lobe and off-axis sources is also observed in NenuFAR observations \citep{munshiHorizonQuantifyingFull2025}.
For both instruments, the target field contributes more power than the off-axis sources, owing to lower beam attenuation. Because of this, the target field dominates the residuals in real data, where calibration and subtraction are less ideal than in these simulations. In our simulations, the only errors that are introduced are beam modelling errors and thermal noise. Because the beam modelling errors are spectrally smooth within the main lobe of the primary beam, we are able to remove the target field sources to a much deeper level, revealing errors in the directions of Cas A and Cyg A.

To test this hypothesis, we also use a combined kernel on the simulation: a 1~MHz kernel is used for subtracting Cas A and Cyg A, whereas a 4~MHz kernel is used for the target field. The results for this dynamic calibration kernel are shown in the final column of \Cref{fig:res_flagging_stats} and are very close to the results for the 1~MHz kernel. This suggests that the outliers found in the data primarily correspond to errors made in the directions of Cas A and Cyg A, such that these dominate the statistics shown in \Cref{fig:res_flagging_stats}.

\Cref{fig:res_flagging_stats} shows two sets of baselines: those between 250~$\lambda$ and 5000~$\lambda$ (i.e. the calibration baselines), and those between 50~$\lambda$ and 250~$\lambda$ (i.e. the EoR analysis baselines). Comparing the calibration and analysis baselines provides information about the level of overfitting, ideally the two sets should be as close as possible. However, there is more residual power on the analysis baselines. This is expected for a baseline cut, and corresponds to earlier findings \citep{mevius_numerical_2021}. Furthermore, the analysis baselines in the standard method require a higher flagging percentage than those in the robust method. 

The horizontal axis coordinates within a panel of \Cref{fig:res_flagging_stats} denote two different models for the off-axis sources, describing them either with a single or multiple point sources. Because the LSTs are identical between the simulations, only the off-axis contribution to the visibilities differs, whereas that of the target field stays the same. Clearly, the single point-source model suffers more from outliers than the multi-point source model, as is evident by the reduction in standard deviation during flagging and the number of flags. This may be due to two things: firstly, the multi-point source model has a bigger spatial footprint, and secondly, each individual component of the multi-point source model contains less power than the single component of the single-point model. To differentiate between these two, we also performed a test with a strongly attenuated single-point model (with the flux reduced by a factor four). This attenuated model suffers from large outliers in the same way as the single point-source model does (see the complete results in \Cref{fig:app_flagging_stats}), such that we can attribute the better calibration behaviour of the multi-point source model to its bigger spatial footprint. As a result of the extended spatial footprint, the source is never fully inside a null, such that its flux is not as abruptly attenuated and the spectral smoothness assumption is not violated as strongly.

The scenario that most closely resembles real LOFAR-EoR observations is the multi-point source model with the 4~MHz regularisation kernel, which was previously found to be preferred over the 1~MHz regularisation kernel \citep{gan_assessing_2023}. The analysis baselines show a clear improvement when using the robust regularisation kernel, reducing both the residual power slightly after flagging from 3.06 to 3.04~Jy and the number of flags from 0.99 to 0.34 per cent on the analysis baselines. Furthermore, the results can potentially be improved even more by utilising the dynamic calibration kernel, which reaches a standard deviation of 3.02~Jy after flagging 0.27 per cent of the simulated data. This is very close to the thermal noise level of 3.01~Jy. However, tests on real data are necessary to confirm that the improved performance of the dynamic calibration kernel is not hampered by real-world effects that are not present in this simulation. 

\subsection{Imaging}
In image space, the main goal of DD-gain calibration of off-axis sources is often to remove them to a sufficient level, so that their PSF-sidelobes do not cause errors in the target field. Although \citet{van_weeren_lofar_2016} have shown that facetting is successful in creating images with long baselines, the power of DD-gain calibration algorithms like {\scriptsize DDECAL} and {\scriptsize SAGECAL-CO} that calibrate in multiple directions simultaneously is that they are also effective in reducing errors on shorter baselines. To measure the effectiveness of DD-gain calibration as a function of baseline length, errors of various angular scales in the image must be probed. 

To this end, the residual data are continuum imaged at a resolution of 0.3 arcmin with a field of $10^\circ\times10^\circ$ and uniform weighting using {\scriptsize WSCLEAN} \citep{offringa_wsclean_2014}. {\scriptsize WSCLEAN} is an efficient wide-field imager, that computes the inverse transform of \Cref{eq:VCZ}. It does so by gridding the visibilities onto a $u\varv$-grid at different levels of the $\varw$-term. Each of these grids are then inverse Fourier transformed to obtain a stack of images. These images are phase shifted with a $\varw$-term correction, and a widefield image is formed by summing the corrected images. \citet{offringa_precision_2019} have shown that this form of imaging is precise enough for 21-cm cosmology. All baselines between 50~$\lambda$ and 5000~$\lambda$ are used in the images. To give a realistic comparison, flags are applied to the data before imaging, such that the outliers are excised. As a result, none of the images contain very bright sources anymore, and the dirty images can be analysed directly, without a need for deconvolution. Subsequently, the images are Fourier transformed to determine their angular power spectrum (i.e. the power at various angular scales),
\begin{equation}
    P(\ell) = \left<\left|\mathcal{F}\left\{I(x,y)\right\}\right|^2\right>_{\ell}.
\end{equation}
Here, $\mathcal{F}\left\{\cdot\right\}$ denotes the spatial Fourier transform, and $I(x,y)$ denotes the image intensity at pixel $(x,y)$. The operator $\left<\cdot\right>_{\ell}$ represents an azimuthal average over Fourier modes with constant angular frequency $\ell$. If $\ell_x$ and $\ell_y$ are the Fourier conjugates of the image axes, $\ell=\sqrt{\ell_x^2+\ell_y^2}$. We use the cosmological convention of using the inverse of the spatial scale as the Fourier conjugate.

\begin{figure}
    \centering
    \includegraphics[width=\linewidth]{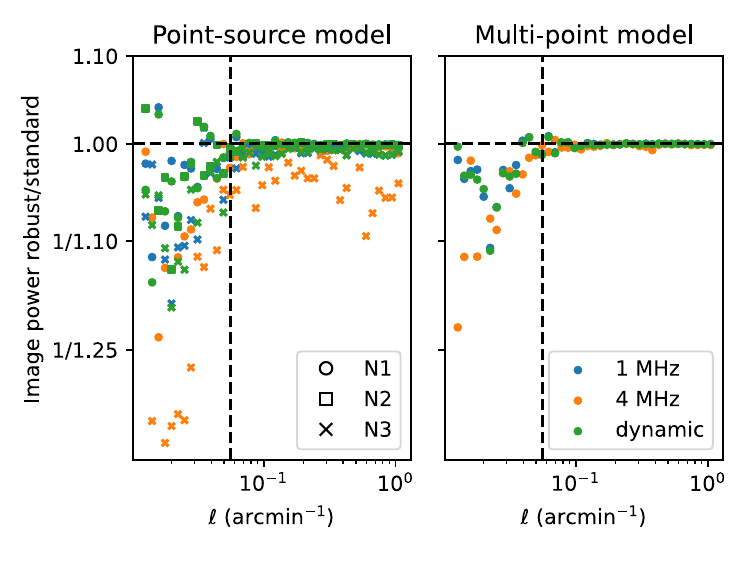}
    \caption{Ratio of power in the residual images with the standard versus proposed method as a function of logarithmically-binned scales in the image. The left panel shows the point-source model for Cas A and Cyg A and the right panel the multi-point model. The colors indicate the different kernels. For the single-point model, the different observing LSTs are denoted with different markers. The horizontal dashed line denotes a ratio of unity, and the vertical dashed line the scale probed by a baseline of $250~\lambda$. }
    \label{fig:res_FT_image}
\end{figure}

If there is residual source power in the images due to gain-calibration errors, this will show as additional power above the thermal noise level. Therefore, better DD-gain calibration generally leads to less power in image space, although sources may leave more residual power on some spatial scales than others. \Cref{fig:res_FT_image} shows the ratio between the power spectrum of the images created using the robust and standard regularisation methods as a function of spatial scale, i.e. $P_{\mathrm{robust}}(\ell)/P_{\mathrm{standard}}(\ell)$, where the subscripts `robust' and `standard' denote the calibration methods. Points below unity generally mean that the robust regularisation method performs better, although the ratio also contains noise. The left panel shows the results with the point-source model (for all nights), and the right panel shows the results with the multi-point model. The different colours indicate different smoothing kernels.

Similar to what was found in visibility space, the robustly regularised DD-gain calibration method performs similarly to or better than the standard method throughout the spectrum, with only a few points exceeding unity, which we attribute to noise. In simulation N3 with the 4~MHz kernel, robust regularisation performs especially well compared to standard regularisation. We expect this to be because the off-axis sources have a higher elevation at the LST range of this simulation. Because of this, they are less attenuated by the antenna beam for this simulation and, therefore, have the highest impact. A similar effect is found for the standard deviation of this simulation in \Cref{app:flags}. This suggests that the robust regularisation method is able to mitigate the impact of strong sources in the sidelobes better on all scales in image space, even when strong spectral regularisation is used.

Overall, however, the ratio remains near unity for both model types, except at the lowest $\ell$-modes, especially those related to scales $\lesssim$0.06~arcmin$^{-1}$. These largest scales in the image correspond to the baselines with a projected length of $\leq250~\lambda$, at which the robust method performs significantly better than the standard method. These scales are important for many science cases, such as 21-cm signal observations of the Epoch of Reionization and Cosmic Dawn, and studies of the diffuse polarised foreground of the Milky Way. We attribute the improvement on the short baselines to the distinction between calibration and analysis baselines. This implies that the robust spectral regulariser results in gains that extrapolate better to shorter baselines than the standard regulariser\footnote{due to the baseline cut, short baseline-gains are constructed from the gain-solutions found on longer baselines.}, which may occur because the gains are less biased by strong spectral variations in the beam-sidelobes. 

\section{Discussion and conclusions} \label{sec:discussion}
In this work, we present simulations of the effect of beam errors due to broken receiver elements in the LOFAR-HBA system that are unaccounted for in state-of-the-art analysis software. We analysed their impact on the calibration of off-axis sources far from the target direction. The beam errors are simulated by switching off antenna elements within LOFAR-HBA tiles at a realistic level of tile degradation (typically up to three receivers per tile of sixteen). We also present a software package, \codename\footnote{\href{https://github.com/Stefanie-B/shimmerr}{https://github.com/Stefanie-B/shimmerr}}, with which similar simulations can be done for other science cases and instruments. We present the resulting beam errors, along with the effects that they have on DD-gain calibration and subtraction of far-field sources. Based on these simulations, and calibration with the state-of-the-art DD-gain calibration code {\scriptsize DDECAL}, we come to the following conclusions:

\vspace{1em}\noindent\textbf{Errors caused by nulls in the calibration beam:} When a bright off-axis source is modelled near a null in the calibration beam, when there is no strong null in the true beam, some of the largest errors in DD-gain calibration solutions occur. Because of the extremely low amplitude and high relative variability of the calibration beam, the required gain corrections in these nulls are typically very large. This is especially the case for arrays with stations with a regular layout. Because the nulls also vary spectrally on scales much smaller than the assumed spectral smoothness scale, the gain solutions in the nulls can cause incorrectly deviating gain solutions to spread to other channels. Nulls in the true beam that are not present in the calibration beam could theoretically cause similar problems, but the effect is smaller because the true beam is smoother than the calibration beam. 

\vspace{1em}\noindent\textbf{Spectral regularisation heuristic:} We show that applying heuristic spectral regularisation, by weighting the spectral smoothing kernel for the gains by the model visibilities, is more robust against the type of errors mentioned above. It outperforms the standard spectral regularisation method in terms of subtracting bright off-axis sources both in visibility and image space, especially on short baselines. Both the amplitude and number of extreme gain-solution failures introduced during calibration are reduced. This holds both before and after outlier excision, for various spectral smoothness scales and off-axis sky-models. Even in the least extreme case investigated, a multi-point source off-axis sky-model, we show that robust regularisation significantly outperforms the standard regularisation method on the shortest baselines. Although we have tested this robust weighting scheme in the DD-gain calibration method {\scriptsize DDECAL} only, it is also straightforward to implement in other DD-gain calibration codes.

We emphasize that this robust reweighting scheme is a simple heuristic that tackles only the incorrect spectral smoothness assumption used in DD-gain calibration. Now that its effectiveness has been shown and a simulation and calibration framework for quick testing is available in \codename, future work can build on this and introduce other regularisation improvements. For example, a more complex basis for modelling errors in a null, that takes into account both spectral and temporal information, can be built. Furthermore, future work can investigate the spatial assumptions in direction-dependent calibration in which a constant gain is assumed for a set of sources (i.e. a calibration cluster). Because direction-dependent beam errors result from variations between elements within a station, they may be better tackled through calibrating the element gains rather than station gains. The spatial structure of the beam then follows from a physically motivated model, which does not require discontinuities.

\vspace{1em}\noindent\textbf{Spectral smoothness scales:} Due to large and small-scale spectral gain changes of the beam far from the target direction, smaller spectral smoothness scales seem to be favoured for off-axis sources than for target field sources. Therefore, using dynamic spectral smoothness scales, where sources that are further from the target direction receive a lower spectral smoothness scale, perform better in the presence of DD beam-modelling errors.

\vspace{1em} The robust regulariser has not been tested in the presence of more general cases of extended emission. This is because current {\scriptsize DDECAL} evaluates the beam response of extended emission at discrete points, effectively removing the difference between point sources and extended sources in the context of beam modelling errors. Furthermore, beam modelling errors are expected to have a lower impact on extended emission, since extended emission does not fully fall into deep beam nulls. Therefore, the difference between the standard and robust regulariser is expected to be smaller for extended emission. Using the difference between the single-point and multi-point off-axis source models as a first-order probe for how the robust regulariser performs for extended emission, we expect extended emission to stabilise the solutions further. 

Diffuse extended emission in the sidelobes of the beam has been shown to pose a challenge for 21-cm cosmology with the drift-scan instrument HERA \citep{n_charles_use_2023}, necessitating a spatial filtering technique to remove this emission prior to calibration. This is not expected to be needed for LOFAR, because LOFAR has more long baselines, allowing it to have a higher baseline cut, that drastically reduces the impact of diffuse emission during gain-calibration. The HERA baseline cut is at 40~m, whereas that of LOFAR is at $250~\lambda\approx 500~$m in the relevant frequency range. Furthermore, \citet{hofer_impact_2025} have shown that the combination of beam errors and diffuse emission does not lead to a bias in the 21-cm power spectrum for LOFAR during direction-independent calibration with a lower baseline cut.

In conclusion, beam modelling for phased-array stations such as those of LOFAR, NenuFAR, MWA (where we consider a tile to be a station) and the upcoming SKA-Low remains a challenging endeavour due to changes in the beam shapes as a function of pointing, time (due to rotation of the Earth), and observing frequency, as well as additional temporal changes due to the behaviour of the telescope due to degrading hardware and weather. The SKA-Low array, for example, will have a higher sensitivity than the current generation of instruments, and likely also have a more complex antenna beam. Therefore, beam effects such as those present for LOFAR-HBA stations will similarly affect SKA-Low. Furthermore, SKA-Low's higher sensitivity may cause far-field sources currently below the noise to appear in the data. Therefore, applying DD-gain calibration methods that are robust against often unavoidable beam-modelling errors in the distant sidelobes and nulls are necessary to use phased-array telescopes to their full capabilities.

\section*{Acknowledgements}
SAB, LVEK, KC, CH, LG, SG, and SM acknowledge the financial support from the European Research Council (ERC) under the European Union’s Horizon 2020 research and innovation programme (Grant agreement No. 884760, “CoDEX”). EC (INAF) would like to acknowledge support from the Centre for Data Science and Systems Complexity (DSSC), Faculty of Science and Engineering at the University of Groningen, and from the Ministry of Universities and Research (MUR) through the PRIN project ‘Optimal inference from radio images of the epoch of reionization’. FGM acknowledges support from the I-DAWN project, funded by the DIM-ORIGINS programme.

Apart from mentioned software that is mentioned in the main body of the text, in this work, we made use of the {\scriptsize DS9} \citep{joye_astronomical_2003} FITS file image viewer, and the {\scriptsize NUMPY} \citep{harris_array_2020}, {\scriptsize NUMBA} \citep{lam_numba_2015}, {\scriptsize ASTROPY} \citep{astropy_collaboration_astropy_2013,astropy_collaboration_astropy_2018,astropy_collaboration_astropy_2022}, {\scriptsize PYTHON-CASACORE} \citep{the_casa_team_casa_2022}, {\scriptsize MATPLOTLIB} \citep{hunter_matplotlib_2007}, and {\scriptsize PANDAS} \citep{mckinney_data_2010} Python packages.

\section*{Data Availability}
The simulation software underlying this article are created with {\scriptsize SHIMMERR}, available at \href{https://doi.org/10.5281/zenodo.15114900}{https://doi.org/10.5281/zenodo.15114900}. The simulations underlying this article will be shared on reasonable request to the corresponding author.



\bibliographystyle{mnras}
\bibliography{references} 

\begin{thebibliography}{}
\makeatletter
\relax
\def\mn@urlcharsother{\let\do\@makeother \do\$\do\&\do\#\do\^\do\_\do\%\do\~}
\def\mn@doi{\begingroup\mn@urlcharsother \@ifnextchar [ {\mn@doi@} {\mn@doi@[]}}
\def\mn@doi@[#1]#2{\def\@tempa{#1}\ifx\@tempa\@empty \href {http://dx.doi.org/#2} {doi:#2}\else \href {http://dx.doi.org/#2} {#1}\fi \endgroup}
\def\mn@eprint#1#2{\mn@eprint@#1:#2::\@nil}
\def\mn@eprint@arXiv#1{\href {http://arxiv.org/abs/#1} {{\tt arXiv:#1}}}
\def\mn@eprint@dblp#1{\href {http://dblp.uni-trier.de/rec/bibtex/#1.xml} {dblp:#1}}
\def\mn@eprint@#1:#2:#3:#4\@nil{\def\@tempa {#1}\def\@tempb {#2}\def\@tempc {#3}\ifx \@tempc \@empty \let \@tempc \@tempb \let \@tempb \@tempa \fi \ifx \@tempb \@empty \def\@tempb {arXiv}\fi \@ifundefined {mn@eprint@\@tempb}{\@tempb:\@tempc}{\expandafter \expandafter \csname mn@eprint@\@tempb\endcsname \expandafter{\@tempc}}}

\bibitem[\protect\citeauthoryear{Albert, van Weeren, Intema  \& Röttgering}{Albert et~al.}{2020}]{albert_probabilistic_2020}
Albert J.~G.,  van Weeren R.~J.,  Intema H.~T.,   Röttgering H. J.~A.,  2020, \mn@doi [Astronomy \& Astrophysics] {10.1051/0004-6361/201937424}, 635, A147

\bibitem[\protect\citeauthoryear{Arras, Frank, Leike, Westermann  \& Enßlin}{Arras et~al.}{2019}]{arras_unified_2019}
Arras P.,  Frank P.,  Leike R.,  Westermann R.,   Enßlin T.~A.,  2019, \mn@doi [Astronomy \& Astrophysics] {10.1051/0004-6361/201935555}, 627, A134

\bibitem[\protect\citeauthoryear{{Astropy Collaboration} et~al.,}{{Astropy Collaboration} et~al.}{2013}]{astropy_collaboration_astropy_2013}
{Astropy Collaboration} et~al., 2013, \mn@doi [Astronomy and Astrophysics] {10.1051/0004-6361/201322068}, 558, A33

\bibitem[\protect\citeauthoryear{{Astropy Collaboration} et~al.,}{{Astropy Collaboration} et~al.}{2018}]{astropy_collaboration_astropy_2018}
{Astropy Collaboration} et~al., 2018, \mn@doi [The Astronomical Journal] {10.3847/1538-3881/aabc4f}, 156, 123

\bibitem[\protect\citeauthoryear{{Astropy Collaboration} et~al.,}{{Astropy Collaboration} et~al.}{2022}]{astropy_collaboration_astropy_2022}
{Astropy Collaboration} et~al., 2022, \mn@doi [The Astrophysical Journal] {10.3847/1538-4357/ac7c74}, 935, 167

\bibitem[\protect\citeauthoryear{Baldwin, Boysen, Hales, Jennings, Waggett, Warner  \& Wilson}{Baldwin et~al.}{1985}]{baldwin_6c_1985}
Baldwin J.~E.,  Boysen R.~C.,  Hales S. E.~G.,  Jennings J.~E.,  Waggett P.~C.,  Warner P.~J.,   Wilson D. M.~A.,  1985, \mn@doi [Monthly Notices of the Royal Astronomical Society] {10.1093/mnras/217.4.717}, 217, 717

\bibitem[\protect\citeauthoryear{Best et~al.,}{Best et~al.}{2023}]{best_lofar_2023}
Best P.~N.,  et~al., 2023, \mn@doi [Monthly Notices of the Royal Astronomical Society] {10.1093/mnras/stad1308}, 523, 1729

\bibitem[\protect\citeauthoryear{Birdi, Repetti  \& Wiaux}{Birdi et~al.}{2020}]{birdi_polca_2020}
Birdi J.,  Repetti A.,   Wiaux Y.,  2020, \mn@doi [Monthly Notices of the Royal Astronomical Society] {10.1093/mnras/stz3555}, 492, 3509

\bibitem[\protect\citeauthoryear{Bowman et~al.,}{Bowman et~al.}{2013}]{bowman_science_2013}
Bowman J.~D.,  et~al., 2013, \mn@doi [Publications of the Astronomical Society of Australia] {10.1017/pas.2013.009}, 30, e031

\bibitem[\protect\citeauthoryear{Brackenhoff et~al.,}{Brackenhoff et~al.}{2024}]{brackenhoff_ionospheric_2024}
Brackenhoff S.~A.,  et~al., 2024, \mn@doi [Monthly Notices of the Royal Astronomical Society] {10.1093/mnras/stae1856}, 533, 632

\bibitem[\protect\citeauthoryear{Ceccotti et~al.,}{Ceccotti et~al.}{2025}]{ceccotti_spectral_2025}
Ceccotti E.,  et~al., 2025, \mn@doi [Astronomy \& Astrophysics] {10.1051/0004-6361/202453106}, 696, A56

\bibitem[\protect\citeauthoryear{Chokshi, Barry, Line, Jordan, Pindor  \& Webster}{Chokshi et~al.}{2024}]{chokshi_necessity_2024}
Chokshi A.,  Barry N.,  Line J. L.~B.,  Jordan C.~H.,  Pindor B.,   Webster R.~L.,  2024, \mn@doi [Monthly Notices of the Royal Astronomical Society] {10.1093/mnras/stae2264}, 534, 2475

\bibitem[\protect\citeauthoryear{Cohen, Lane, Cotton, Kassim, Lazio, Perley, Condon  \& Erickson}{Cohen et~al.}{2007}]{cohen_vla_2007}
Cohen A.~S.,  Lane W.~M.,  Cotton W.~D.,  Kassim N.~E.,  Lazio T. J.~W.,  Perley R.~A.,  Condon J.~J.,   Erickson W.~C.,  2007, \mn@doi [The Astronomical Journal] {10.1086/520719}, 134, 1245

\bibitem[\protect\citeauthoryear{Condon, Cotton, Greisen, Yin, Perley, Taylor  \& Broderick}{Condon et~al.}{1998}]{condon_nrao_1998}
Condon J.~J.,  Cotton W.~D.,  Greisen E.~W.,  Yin Q.~F.,  Perley R.~A.,  Taylor G.~B.,   Broderick J.~J.,  1998, \mn@doi [The Astronomical Journal] {10.1086/300337}, 115, 1693

\bibitem[\protect\citeauthoryear{Cronyn}{Cronyn}{1972}]{cronyn_interferometer_1972}
Cronyn W.~M.,  1972, \mn@doi [The Astrophysical Journal] {10.1086/151480}, 174, 181

\bibitem[\protect\citeauthoryear{Dewdney, Hall, Schilizzi  \& Lazio}{Dewdney et~al.}{2009}]{dewdney_square_2009}
Dewdney P.~E.,  Hall P.~J.,  Schilizzi R.~T.,   Lazio T. J. L.~W.,  2009, \mn@doi [Proceedings of the IEEE] {10.1109/JPROC.2009.2021005}, 97, 1482

\bibitem[\protect\citeauthoryear{Edge, Shakeshaft, McAdam, Baldwin  \& Archer}{Edge et~al.}{1959}]{edge_survey_1959}
Edge D.~O.,  Shakeshaft J.~R.,  McAdam W.~B.,  Baldwin J.~E.,   Archer S.,  1959, Memoirs of the Royal Astronomical Society, 68, 37

\bibitem[\protect\citeauthoryear{Elder \& Jacobs}{Elder \& Jacobs}{2024}]{elder_investigating_2024}
Elder K.,  Jacobs D.~C.,  2024, Investigating mutual coupling in the {MWA} {Phase} {II} compact array, \mn@doi{10.48550/arXiv.2411.04193}, \url {http://arxiv.org/abs/2411.04193}

\bibitem[\protect\citeauthoryear{Ewall-Wice, Dillon, Liu  \& Hewitt}{Ewall-Wice et~al.}{2017}]{ewall-wice_impact_2017}
Ewall-Wice A.,  Dillon J.~S.,  Liu A.,   Hewitt J.,  2017, \mn@doi [Monthly Notices of the Royal Astronomical Society] {10.1093/mnras/stx1221}, 470, 1849

\bibitem[\protect\citeauthoryear{Franzen et~al.,}{Franzen et~al.}{2016}]{franzen_154_2016}
Franzen T. M.~O.,  et~al., 2016, \mn@doi [Monthly Notices of the Royal Astronomical Society] {10.1093/mnras/stw823}, 459, 3314

\bibitem[\protect\citeauthoryear{Gan et~al.,}{Gan et~al.}{2022}]{gan_statistical_2022}
Gan H.,  et~al., 2022, \mn@doi [Astronomy \& Astrophysics] {10.1051/0004-6361/202142945}, 663, A9

\bibitem[\protect\citeauthoryear{Gan et~al.,}{Gan et~al.}{2023}]{gan_assessing_2023}
Gan H.,  et~al., 2023, \mn@doi [Astronomy \& Astrophysics] {10.1051/0004-6361/202244316}, 669, A20

\bibitem[\protect\citeauthoryear{Hale et~al.,}{Hale et~al.}{2019}]{hale_lofar_2019}
Hale C.~L.,  et~al., 2019, \mn@doi [Astronomy \& Astrophysics] {10.1051/0004-6361/201833906}, 622, A4

\bibitem[\protect\citeauthoryear{Hales, Riley, Waldram, Warner  \& Baldwin}{Hales et~al.}{2007}]{hales_final_2007}
Hales S. E.~G.,  Riley J.~M.,  Waldram E.~M.,  Warner P.~J.,   Baldwin J.~E.,  2007, \mn@doi [Monthly Notices of the Royal Astronomical Society] {10.1111/j.1365-2966.2007.12392.x}, 382, 1639

\bibitem[\protect\citeauthoryear{Hamaker, Bregman  \& Sault}{Hamaker et~al.}{1996}]{hamaker_understanding_1996}
Hamaker J.~P.,  Bregman J.~D.,   Sault R.~J.,  1996, \mn@doi [Astronomy and Astrophysics Supplement Series] {10.1051/aas:1996146}, 117, 137

\bibitem[\protect\citeauthoryear{Harris et~al.,}{Harris et~al.}{2020}]{harris_array_2020}
Harris C.~R.,  et~al., 2020, \mn@doi [Nature] {10.1038/s41586-020-2649-2}, 585, 357

\bibitem[\protect\citeauthoryear{Heald, McKean  \& Pizzo}{Heald et~al.}{2018}]{heald_low_2018}
Heald G.,  McKean J.,   Pizzo R.,  eds, 2018, Low {Frequency} {Radio} {Astronomy} and the {LOFAR} {Observatory}: {Lectures} from the {Third} {LOFAR} {Data} {Processing} {School}.
 Astrophysics and {Space} {Science} {Library} Vol. 426, Springer International Publishing, Cham, \mn@doi{10.1007/978-3-319-23434-2}, \url {http://link.springer.com/10.1007/978-3-319-23434-2}

\bibitem[\protect\citeauthoryear{Hunter}{Hunter}{2007}]{hunter_matplotlib_2007}
Hunter J.~D.,  2007, \mn@doi [Computing in Science \& Engineering] {10.1109/MCSE.2007.55}, 9, 90

\bibitem[\protect\citeauthoryear{Hurley-Walker et~al.,}{Hurley-Walker et~al.}{2017}]{hurley-walker_galactic_2017}
Hurley-Walker N.,  et~al., 2017, \mn@doi [Monthly Notices of the Royal Astronomical Society] {10.1093/mnras/stw2337}, 464, 1146

\bibitem[\protect\citeauthoryear{Höfer et~al.,}{Höfer et~al.}{2025}]{hofer_impact_2025}
Höfer C.,  et~al., 2025, The impact of diffuse {Galactic} emission on direction-independent gain calibration in high-redshift 21 cm observations, \mn@doi{10.48550/arXiv.2504.03554}, \url {http://arxiv.org/abs/2504.03554}

\bibitem[\protect\citeauthoryear{Iheanetu, Girard, Smirnov, Asad, de Villiers, Thorat, Makhathini  \& Perley}{Iheanetu et~al.}{2019}]{iheanetu_primary_2019}
Iheanetu K.,  Girard J.~N.,  Smirnov O.,  Asad K. M.~B.,  de Villiers M.,  Thorat K.,  Makhathini S.,   Perley R.~A.,  2019, \mn@doi [Monthly Notices of the Royal Astronomical Society] {10.1093/mnras/stz702}, 485, 4107

\bibitem[\protect\citeauthoryear{Intema, Jagannathan, Mooley  \& Frail}{Intema et~al.}{2017}]{intema_gmrt_2017}
Intema H.~T.,  Jagannathan P.,  Mooley K.~P.,   Frail D.~A.,  2017, \mn@doi [Astronomy and Astrophysics] {10.1051/0004-6361/201628536}, 598, A78

\bibitem[\protect\citeauthoryear{Jacobs et~al.,}{Jacobs et~al.}{2017}]{jacobs_first_2017}
Jacobs D.~C.,  et~al., 2017, \mn@doi [Publications of the Astronomical Society of the Pacific] {10.1088/1538-3873/aa56b9}, 129, 035002

\bibitem[\protect\citeauthoryear{Joye, Mandel, Payne, Jedrzejewski  \& Hook}{Joye et~al.}{2003}]{joye_astronomical_2003}
Joye W.~A.,  Mandel E.,  Payne H.~E.,  Jedrzejewski R.~I.,   Hook R.~N.,  2003, in Astronomical {Society} of the {Pacific} {Conference} {Series}, ed. {HE} {Payne}, {RI} {Jedrzejewski}, \& {RN} {Hook}.

\bibitem[\protect\citeauthoryear{Koopmans}{Koopmans}{2010}]{koopmans_ionospheric_2010}
Koopmans L. V.~E.,  2010, \mn@doi [The Astrophysical Journal] {10.1088/0004-637X/718/2/963}, 718, 963

\bibitem[\protect\citeauthoryear{Koopmans et~al.,}{Koopmans et~al.}{2015}]{koopmans_cosmic_2015}
Koopmans L. V.~E.,  et~al., 2015, in Proceedings of {Advancing} {Astrophysics} with the {Square} {Kilometre} {Array} — {PoS}({AASKA14}). p.~001, \mn@doi{10.22323/1.215.0001}

\bibitem[\protect\citeauthoryear{Lam, Pitrou  \& Seibert}{Lam et~al.}{2015}]{lam_numba_2015}
Lam S.~K.,  Pitrou A.,   Seibert S.,  2015, in Proceedings of the {Second} {Workshop} on the {LLVM} {Compiler} {Infrastructure} in {HPC}. {LLVM} '15.
Association for Computing Machinery, New York, NY, USA, pp~1--6, \mn@doi{10.1145/2833157.2833162}, \url {https://dl.acm.org/doi/10.1145/2833157.2833162}

\bibitem[\protect\citeauthoryear{Line et~al.,}{Line et~al.}{2018}]{line_situ_2018}
Line J. L.~B.,  et~al., 2018, \mn@doi [Publications of the Astronomical Society of Australia] {10.1017/pasa.2018.30}, 35, e045

\bibitem[\protect\citeauthoryear{Lonsdale}{Lonsdale}{2005}]{lonsdale_configuration_2005}
Lonsdale C.~J.,  2005, in From {Clark} {Lake} to the {Long} {Wavelength} {Array}: {Bill} {Erickson}'s {Radio} {Science}. p.~399

\bibitem[\protect\citeauthoryear{McKinney}{McKinney}{2010}]{mckinney_data_2010}
McKinney W.,  2010, SciPy, 445, 51

\bibitem[\protect\citeauthoryear{Mertens et~al.,}{Mertens et~al.}{2020}]{mertens_improved_2020}
Mertens F.~G.,  et~al., 2020, \mn@doi [Monthly Notices of the Royal Astronomical Society] {10.1093/mnras/staa327}, 493, 1662

\bibitem[\protect\citeauthoryear{Mertens et~al.,}{Mertens et~al.}{2025}]{mertens_deeper_2025}
Mertens F.~G.,  et~al., 2025, \mn@doi [Astronomy \& Astrophysics] {10.1051/0004-6361/202554158}

\bibitem[\protect\citeauthoryear{Mevius et~al.,}{Mevius et~al.}{2021}]{mevius_numerical_2021}
Mevius M.,  et~al., 2021, \mn@doi [Monthly Notices of the Royal Astronomical Society] {10.1093/mnras/stab3233}, 509, 3693

\bibitem[\protect\citeauthoryear{Mouri Sardarabadi \& Koopmans}{Mouri Sardarabadi \& Koopmans}{2019}]{mourisardarabadi_quantifying_2019}
Mouri Sardarabadi A.,  Koopmans L. V.~E.,  2019, \mn@doi [Monthly Notices of the Royal Astronomical Society] {10.1093/mnras/sty3444}, 483, 5480

\bibitem[\protect\citeauthoryear{Munshi et~al.,}{Munshi et~al.}{2024}]{munshi_first_2024}
Munshi S.,  et~al., 2024, \mn@doi [Astronomy \& Astrophysics] {10.1051/0004-6361/202348329}, 681, A62

\bibitem[\protect\citeauthoryear{Munshi et~al.,}{Munshi et~al.}{2025}]{munshiHorizonQuantifyingFull2025}
Munshi S.,  et~al., 2025, \mn@doi [Astronomy \& Astrophysics] {10.1051/0004-6361/202451181}, 693, A276

\bibitem[\protect\citeauthoryear{Murphy et~al.,}{Murphy et~al.}{2013}]{murphy_vast_2013}
Murphy T.,  et~al., 2013, \mn@doi [Publications of the Astronomical Society of Australia] {10.1017/pasa.2012.006}, 30, e006

\bibitem[\protect\citeauthoryear{{N. Charles}, {N. Kern}, {G. Bernardi}, {L. Bester}, {O. Smirnov}, {N. Fagnoni}  \& {E. Acedo}}{{N. Charles} et~al.}{2023}]{n_charles_use_2023}
{N. Charles} {N. Kern} {G. Bernardi} {L. Bester} {O. Smirnov} {N. Fagnoni}  {E. Acedo} 2023, \mn@doi [Monthly notices of the Royal Astronomical Society] {10.1093/mnras/stad1046}

\bibitem[\protect\citeauthoryear{Nunhokee et~al.,}{Nunhokee et~al.}{2020}]{nunhokee_measuring_2020}
Nunhokee C.~D.,  et~al., 2020, \mn@doi [The Astrophysical Journal] {10.3847/1538-4357/ab9634}, 897, 5

\bibitem[\protect\citeauthoryear{O'Hara et~al.,}{O'Hara et~al.}{2024}]{ohara_uncovering_2024}
O'Hara O. S.~D.,  et~al., 2024, Uncovering the {Effects} of {Array} {Mutual} {Coupling} in 21-cm {Experiments} with the {SKA}-{Low} {Radio} {Telescope}, \mn@doi{10.48550/arXiv.2412.01699}, \url {http://arxiv.org/abs/2412.01699}

\bibitem[\protect\citeauthoryear{Offringa, van~de Gronde  \& Roerdink}{Offringa et~al.}{2012}]{offringa_morphological_2012}
Offringa A.~R.,  van~de Gronde J.~J.,   Roerdink J. B. T.~M.,  2012, \mn@doi [Astronomy \& Astrophysics] {10.1051/0004-6361/201118497}, 539, A95

\bibitem[\protect\citeauthoryear{Offringa et~al.,}{Offringa et~al.}{2014}]{offringa_wsclean_2014}
Offringa A.~R.,  et~al., 2014, \mn@doi [Monthly Notices of the Royal Astronomical Society] {10.1093/mnras/stu1368}, 444, 606

\bibitem[\protect\citeauthoryear{Offringa, Mertens, Tol, Veenboer, Gehlot, Koopmans  \& Mevius}{Offringa et~al.}{2019}]{offringa_precision_2019}
Offringa A.~R.,  Mertens F.,  Tol S. v.~d.,  Veenboer B.,  Gehlot B.~K.,  Koopmans L. V.~E.,   Mevius M.,  2019, \mn@doi [Astronomy \& Astrophysics] {10.1051/0004-6361/201935722}, 631, A12

\bibitem[\protect\citeauthoryear{Patil et~al.,}{Patil et~al.}{2016}]{patil_systematic_2016}
Patil A.~H.,  et~al., 2016, \mn@doi [Monthly Notices of the Royal Astronomical Society] {10.1093/mnras/stw2277}, 463, 4317

\bibitem[\protect\citeauthoryear{Patil et~al.,}{Patil et~al.}{2017}]{patil_upper_2017}
Patil A.~H.,  et~al., 2017, \mn@doi [The Astrophysical Journal] {10.3847/1538-4357/aa63e7}, 838, 65

\bibitem[\protect\citeauthoryear{Pilkington \& Scott}{Pilkington \& Scott}{1965}]{pilkington_survey_1965}
Pilkington J. D.~H.,  Scott J.~F.,  1965, Memoirs of the Royal Astronomical Society, 69, 183

\bibitem[\protect\citeauthoryear{Rengelink, Tang, de Bruyn, Miley, Bremer, Roettgering  \& Bremer}{Rengelink et~al.}{1997}]{rengelink_westerbork_1997}
Rengelink R.~B.,  Tang Y.,  de Bruyn A.~G.,  Miley G.~K.,  Bremer M.~N.,  Roettgering H. J.~A.,   Bremer M. A.~R.,  1997, \mn@doi [Astronomy and Astrophysics Supplement Series] {10.1051/aas:1997358}, 124, 259

\bibitem[\protect\citeauthoryear{Repetti, Birdi, Dabbech  \& Wiaux}{Repetti et~al.}{2017}]{repetti_non-convex_2017}
Repetti A.,  Birdi J.,  Dabbech A.,   Wiaux Y.,  2017, \mn@doi [Monthly Notices of the Royal Astronomical Society] {10.1093/mnras/stx1267}, 470, 3981

\bibitem[\protect\citeauthoryear{Roth, Arras, Reinecke, Perley, Westermann  \& Enßlin}{Roth et~al.}{2023}]{roth_bayesian_2023}
Roth J.,  Arras P.,  Reinecke M.,  Perley R.~A.,  Westermann R.,   Enßlin T.~A.,  2023, \mn@doi [Astronomy \& Astrophysics] {10.1051/0004-6361/202346851}, 678, A177

\bibitem[\protect\citeauthoryear{Sardarabadi \& Koopmans}{Sardarabadi \& Koopmans}{2019}]{sardarabadi_identifiability_2019}
Sardarabadi A.~M.,  Koopmans L. V.~E.,  2019, arXiv:1902.02482 [astro-ph]

\bibitem[\protect\citeauthoryear{Shimwell et~al.,}{Shimwell et~al.}{2017}]{shimwell_lofar_2017}
Shimwell T.~W.,  et~al., 2017, \mn@doi [Astronomy \& Astrophysics] {10.1051/0004-6361/201629313}, 598, A104

\bibitem[\protect\citeauthoryear{Shimwell et~al.,}{Shimwell et~al.}{2022}]{shimwell_lofar_2022}
Shimwell T.~W.,  et~al., 2022, \mn@doi [Astronomy \& Astrophysics] {10.1051/0004-6361/202142484}, 659, A1

\bibitem[\protect\citeauthoryear{Smirnov}{Smirnov}{2011a}]{smirnov_revisiting_2011-1}
Smirnov O.~M.,  2011a, \mn@doi [Astronomy \& Astrophysics] {10.1051/0004-6361/201016082}, 527, A106

\bibitem[\protect\citeauthoryear{Smirnov}{Smirnov}{2011b}]{smirnov_revisiting_2011}
Smirnov O.~M.,  2011b, \mn@doi [Astronomy \& Astrophysics] {10.1051/0004-6361/201116434}, 527, A107

\bibitem[\protect\citeauthoryear{Smirnov \& Tasse}{Smirnov \& Tasse}{2015}]{smirnov_radio_2015}
Smirnov O.~M.,  Tasse C.,  2015, \mn@doi [Monthly Notices of the Royal Astronomical Society] {10.1093/mnras/stv418}, 449, 2668

\bibitem[\protect\citeauthoryear{The CASA~Team}{The CASA~Team}{2022}]{the_casa_team_casa_2022}
The CASA~Team e.~a.,  2022, \mn@doi [Publications of the Astronomical Society of the Pacific] {10.1088/1538-3873/ac9642}, 134, 114501

\bibitem[\protect\citeauthoryear{Thompson, Moran  \& {George W. Swenson Jr}}{Thompson et~al.}{2001}]{thompson_interferometry_2001}
Thompson A.~R.,  Moran J.~M.,   {George W. Swenson Jr} 2001, Interferometry and {Synthesis} in {Radio} {Astronomy}.
Springer Nature

\bibitem[\protect\citeauthoryear{Trott}{Trott}{2021}]{trott_impact_2021}
Trott C.~M.,  2021, \mn@doi [Journal of Astronomical Telescopes, Instruments, and Systems] {10.1117/1.JATIS.8.1.011011}, 8, 011011

\bibitem[\protect\citeauthoryear{Trott \& Wayth}{Trott \& Wayth}{2016}]{trott_spectral_2016}
Trott C.~M.,  Wayth R.~B.,  2016, \mn@doi [Publications of the Astronomical Society of Australia] {10.1017/pasa.2016.18}, 33, e019

\bibitem[\protect\citeauthoryear{Trott et~al.,}{Trott et~al.}{2020}]{trott_deep_2020}
Trott C.~M.,  et~al., 2020, \mn@doi [Monthly Notices of the Royal Astronomical Society] {10.1093/mnras/staa414}, 493, 4711

\bibitem[\protect\citeauthoryear{Wijnholds, Bregman  \& van Ardenne}{Wijnholds et~al.}{2011}]{wijnholds_calibratability_2011}
Wijnholds S.~J.,  Bregman J.~D.,   van Ardenne A.,  2011, \mn@doi [Radio Science] {10.1029/2011RS004733}, 46

\bibitem[\protect\citeauthoryear{Wijnholds, Arts, Bolli, di Ninni  \& Virone}{Wijnholds et~al.}{2019}]{wijnholds_using_2019}
Wijnholds S.~J.,  Arts M.,  Bolli P.,  di Ninni P.,   Virone G.,  2019, in 2019 {International} {Conference} on {Electromagnetics} in {Advanced} {Applications} ({ICEAA}). pp 0437--0442, \mn@doi{10.1109/ICEAA.2019.8878949}, \url {https://ieeexplore.ieee.org/abstract/document/8878949}

\bibitem[\protect\citeauthoryear{Wilensky et~al.,}{Wilensky et~al.}{2024}]{wilensky_high-dimensional_2024}
Wilensky M.~J.,  et~al., 2024, \mn@doi [RAS Techniques and Instruments] {10.1093/rasti/rzae029}, 3, 400

\bibitem[\protect\citeauthoryear{Yatawatta}{Yatawatta}{2015}]{yatawatta_distributed_2015}
Yatawatta S.,  2015, \mn@doi [Monthly Notices of the Royal Astronomical Society] {10.1093/mnras/stv596}, 449, 4506

\bibitem[\protect\citeauthoryear{Zarka, Girard, Tagger  \& Denis}{Zarka et~al.}{2012}]{zarka_lssnenufar_2012}
Zarka P.,  Girard J.~N.,  Tagger M.,   Denis L.,  2012, in {SF2A}-2012: {Proceedings} of the {Annual} meeting of the {French} {Society} of {Astronomy} and {Astrophysics}. pp 687--694

\bibitem[\protect\citeauthoryear{de Gasperin et~al.,}{de~Gasperin et~al.}{2019}]{de_gasperin_systematic_2019}
de Gasperin F.,  et~al., 2019, \mn@doi [Astronomy \& Astrophysics] {10.1051/0004-6361/201833867}, 622, A5

\bibitem[\protect\citeauthoryear{de Gasperin et~al.,}{de~Gasperin et~al.}{2023}]{de_gasperin_lofar_2023}
de Gasperin F.,  et~al., 2023, \mn@doi [Astronomy \& Astrophysics] {10.1051/0004-6361/202245389}, 673, A165

\bibitem[\protect\citeauthoryear{van Diepen, Dijkema  \& Offringa}{van Diepen et~al.}{2018}]{van_diepen_dppp_2018}
van Diepen G.,  Dijkema T.~J.,   Offringa A.,  2018, Astrophysics Source Code Library, p. ascl:1804.003

\bibitem[\protect\citeauthoryear{van Haarlem et~al.,}{van Haarlem et~al.}{2013}]{van_haarlem_lofar_2013}
van Haarlem M.~P.,  et~al., 2013, \mn@doi [Astronomy \& Astrophysics] {10.1051/0004-6361/201220873}, 556, A2

\bibitem[\protect\citeauthoryear{van Weeren et~al.,}{van Weeren et~al.}{2016}]{van_weeren_lofar_2016}
van Weeren R.~J.,  et~al., 2016, \mn@doi [The Astrophysical Journal Supplement Series] {10.3847/0067-0049/223/1/2}, 223, 2

\bibitem[\protect\citeauthoryear{van~der Tol, Jeffs  \& van~der Veen}{van~der Tol et~al.}{2007}]{van_der_tol_self-calibration_2007}
van~der Tol S.,  Jeffs B.~D.,   van~der Veen A.-J.,  2007, \mn@doi [IEEE Transactions on Signal Processing] {10.1109/TSP.2007.896243}, 55, 4497

\makeatother
\end{thebibliography}




\appendix

\section{Residual visibility statistics} \label{app:flags}
\begin{figure*}
    \centering
    \includegraphics[width=\linewidth]{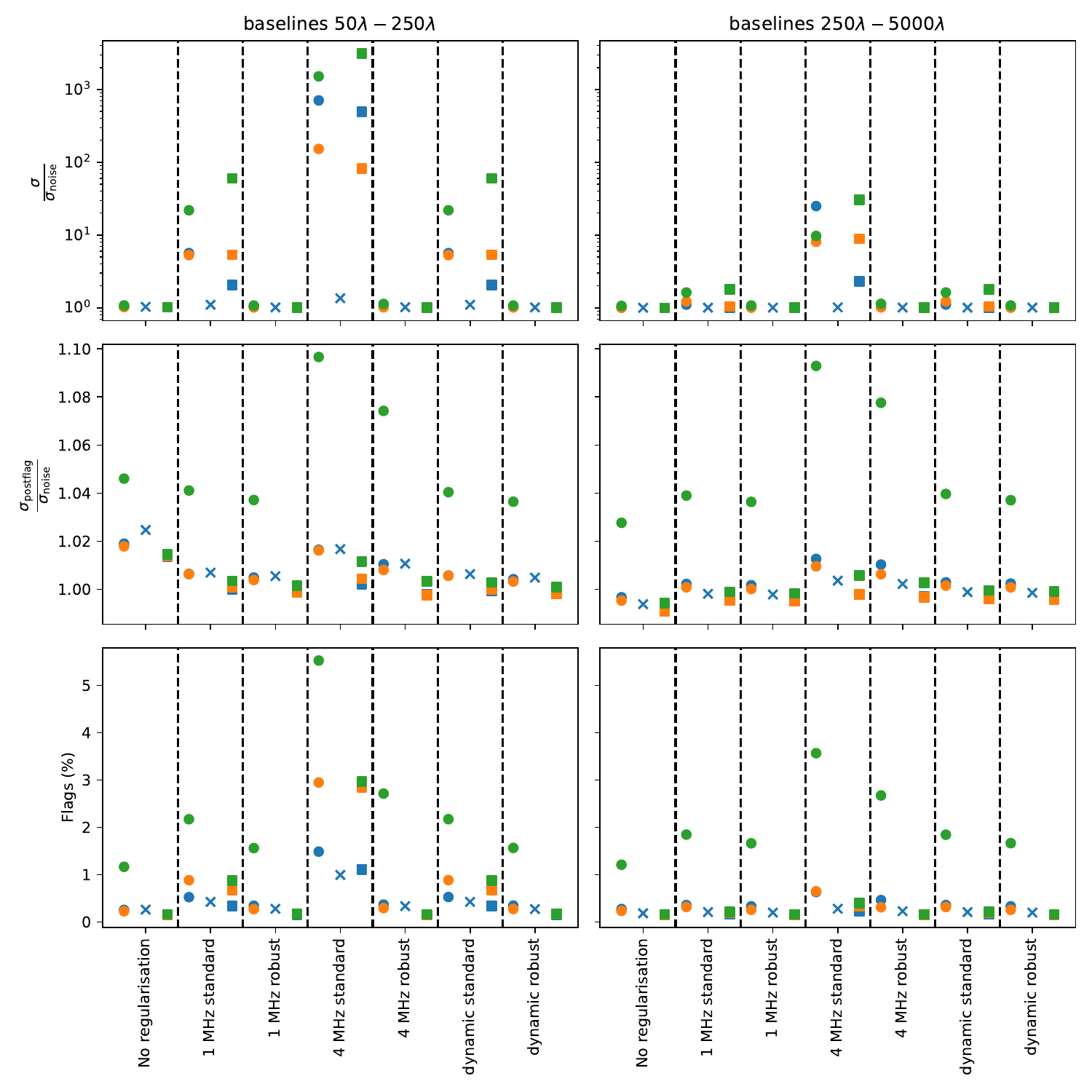}
    \caption{Standard deviations compared to the noise level before and after flagging and flagging percentages for all simulations under consideration in this work. We split the statistics into analysis baselines ($<250~\lambda$, left-hand column) and calibration baselines ($>250~\lambda$, right-hand column). The circles, crosses and squares denote the single point-source, multi-point source and attenuated models for Cas A and Cyg A, respectively, and are horizontally offset for legibility. The colors denote different observing LSTs, with blue, orange, and green representing N1, N2, and N3, respectively. The bins within the panels, separated with dashed lines, show the different regularisation settings using in calibration.}
    \label{fig:app_flagging_stats}
\end{figure*}

In \Cref{fig:res_flagging_stats}, the standard deviations and flagging fractions of two of the simulations in this work are shown. \Cref{fig:app_flagging_stats} in this appendix contains these statistics for all seven simulations, in all considered calibration setups.

\bsp	
\label{lastpage}
\end{document}